# Electron Fourier ptychography for phase reconstruction

Jingjing Zhao[1*], Chen Huang[1], Ali Mostaed[1], Amirafshar Moshtaghpour[1], James M. Parkhurst[1,2], Ivan Lobato[1], Marcus Gallagher-Jones[1], Judy S. Kim[1,3], Mark Boyce[4], David Stuart[4], Elena A. Andreeva[5], Jacques-Philippe Colletier[6], Angus I. Kirkland[1,3]

[1]The Rosalind Franklin Institute, Didcot, OX11 0QX, UK.
[2]Diamond Light Source, Didcot, OX11 0DE, UK.
[3]Department of Materials, University of Oxford, Oxford, OX1 3PH, UK.
[4]Division of Structural Biology, Wellcome Centre for Human Genetics, University of Oxford, Oxford, OX3 7BN, UK.
[5]Max Planck Institute for Medical Research, Heidelberg, D-69120, Germany.
[6]Univ. Grenoble Alpes, CNRS, CEA, Institut de Biologie Structurale, Grenoble, F-38000, France.

[*]To whom correspondence may be addressed.
Email: jingjing.zhao@rfi.ac.uk

## Abstract

Exit wavefunction reconstruction is important in transmission electron microscopy for structural studies. We describe electron Fourier ptychography and its application to phase reconstruction of both radiation-resistant and beam-sensitive materials. We demonstrate that the phase of the exit wave can be reconstructed to high resolution using a modified iterative phase retrieval algorithm from data collected in an alternative optical geometry. This method achieves a spatial resolution of 0.63 nm at a fluence of $4.5\times10^2$ $e^-/nm^2$, as validated on Cry11Aa protein crystals under cryogenic conditions. Notably, this method requires no instrumental modifications, is straightforward to implement, and can be seamlessly integrated with existing data collection software, providing a broadly accessible alternative approach for structural studies.

## Introduction

Efficient reconstruction of the phase of the specimen exit wave is important in many aspects of electron microscopy. Ptychography, as one approach, was originally described by Hoppe and Hegerl[1] and has been successfully implemented using various radiations[2]. Notably, electron ptychography has been used for structure determination in materials science[3–13] and structural biology[14–19] driven in part by the development of fast direct electron detectors[20,21]. Data acquisition for conventional electron ptychography is based on scanning transmission electron microscopy (STEM), in which the sample is scanned by a convergent focussed[9] or defocussed probe[17] with an array of diffraction patterns collected in the far field. As an alternative, Fourier ptychography scans Fourier space by tilting plane-wave illumination[22–26], mechanical scanning of a confined objective aperture[27–29], or moving cameras[27,30,31], all of which generate a dataset of real-space images. This approach, which originated from techniques used in astronomy[32] and radar[33], has been used with visible light[34–37] and X-rays[38,39]. At optical wavelengths, Fourier ptychography allows wide field reconstruction at resolution beyond the optical instrumental limit[36,37], and has been used to record high-speed videos for *in-vitro* studies[40,41]. In electron microscopy, Kirkland, Saxton, and co-workers implemented this method using tilted illumination in the 1980s[22–24], and demonstrated reconstruction of the complex exit wave based on an analytical linear restoration[23,42] at a resolution higher than the optical axial limit of an uncorrected instrument.

In this work, we describe an implementation of electron Fourier ptychography (eFP) for both spherical aberration ($C_3$) corrected and uncorrected microscopes, using a modified Ptychographic Iterative Engine (PIE)[43,44]. We further show that the specimen exit wave can be reconstructed at high resolution from both a radiation-resistant sample and demonstrate applications to low fluence exit wave reconstruction of vitrified biological specimens.

## Results

### Implementation

In the data acquisition geometry used (**Fig. 1**), plane-wave illumination was tilted to several defined incident angles and azimuths such that the information transferred at each tilt overlapped in Fourier space. Tilting the illumination effectively shifts the objective lens transfer function from the optical axis and consequently, higher resolution information is transferred in one direction beyond the axial information transfer limit[22–24,42]. Several real space images recorded at different azimuths can subsequently be used to reconstruct the specimen exit wave with rotationally symmetric transfer. This experimental data acquisition is similar to that reported previously[22–24,42] but does not include a focal series of images at each tilt (from which the axial aberrations can be fitted[45,46]) in order to minimise the overall electron fluence.

The underlying theory of tilted illumination imaging has been described previously[22–24,47] and is only summarised here for convenience.

For tilted plane-wave illumination, the complex valued incident wave $\psi_{inc}(\boldsymbol{r}, \boldsymbol{k_\tau})$ can be written as:

$$\psi_{inc}(\boldsymbol{r}, \boldsymbol{k_\tau}) = \exp(2\pi i \boldsymbol{k_\tau} \cdot \boldsymbol{r}) \quad (1)$$

where $\boldsymbol{k_\tau}$ is the two-dimensional wavevector of the tilted incident beam, defined in terms of the beam tilt $\boldsymbol{\tau}$ as $\boldsymbol{k_\tau} = \boldsymbol{\tau}/\lambda$ and with the vector $\boldsymbol{r}$ representing a two-dimensional position in real space.

The specimen exit wave is given by:

$$\psi_{ex}(\boldsymbol{r}, \boldsymbol{k_\tau}) = O(\boldsymbol{r})\exp(2\pi i \boldsymbol{k_\tau} \cdot \boldsymbol{r}) \quad (2)$$

where $O(\boldsymbol{r})$ is the transmission function of a thin object in two dimensions.

The image wave in Fourier space is thus:

$$\Psi_{im}(\boldsymbol{k}, \boldsymbol{k_\tau}) = \mathcal{F}[O(\boldsymbol{r}) \exp(2\pi i \boldsymbol{k_\tau} \cdot \boldsymbol{r})]w(\boldsymbol{k}) = \Psi_{ex}(\boldsymbol{k}, \boldsymbol{k_\tau})w(\boldsymbol{k}) \quad (3)$$

where $w(\boldsymbol{k})$ is the wave transfer function and $\boldsymbol{k}$ is a two-dimensional vector in reciprocal space. For a sample where the weak phase object approximation is valid, the tilt angle can be included in the wave transfer function[48] and Eq. (3) can be rewritten as:

$$\Psi_{im}(\boldsymbol{k}, \boldsymbol{k_\tau}) = \Psi_{ex}(\boldsymbol{k})w'(\boldsymbol{k}, \boldsymbol{k_\tau}) \quad (4)$$

in which $w'(\boldsymbol{k}, \boldsymbol{k_\tau})$ represents an effective wave transfer function, with details provided in Supplementary **Text S1**. Previously reported work[22–24,42,47] has used this formulation in which the beam tilt is incorporated into the wave transfer function, and the exit wave $\Psi_{ex}(\boldsymbol{k})$ is unchanged for tilted illumination. However, in this work, we have used the formulation defined by Eq. (3) which does not require the weak phase object approximation to be satisfied. However, the effective wave transfer function[48] (Eq. (4)) was used for analysing optimal beam tilt angles for different experimental conditions.

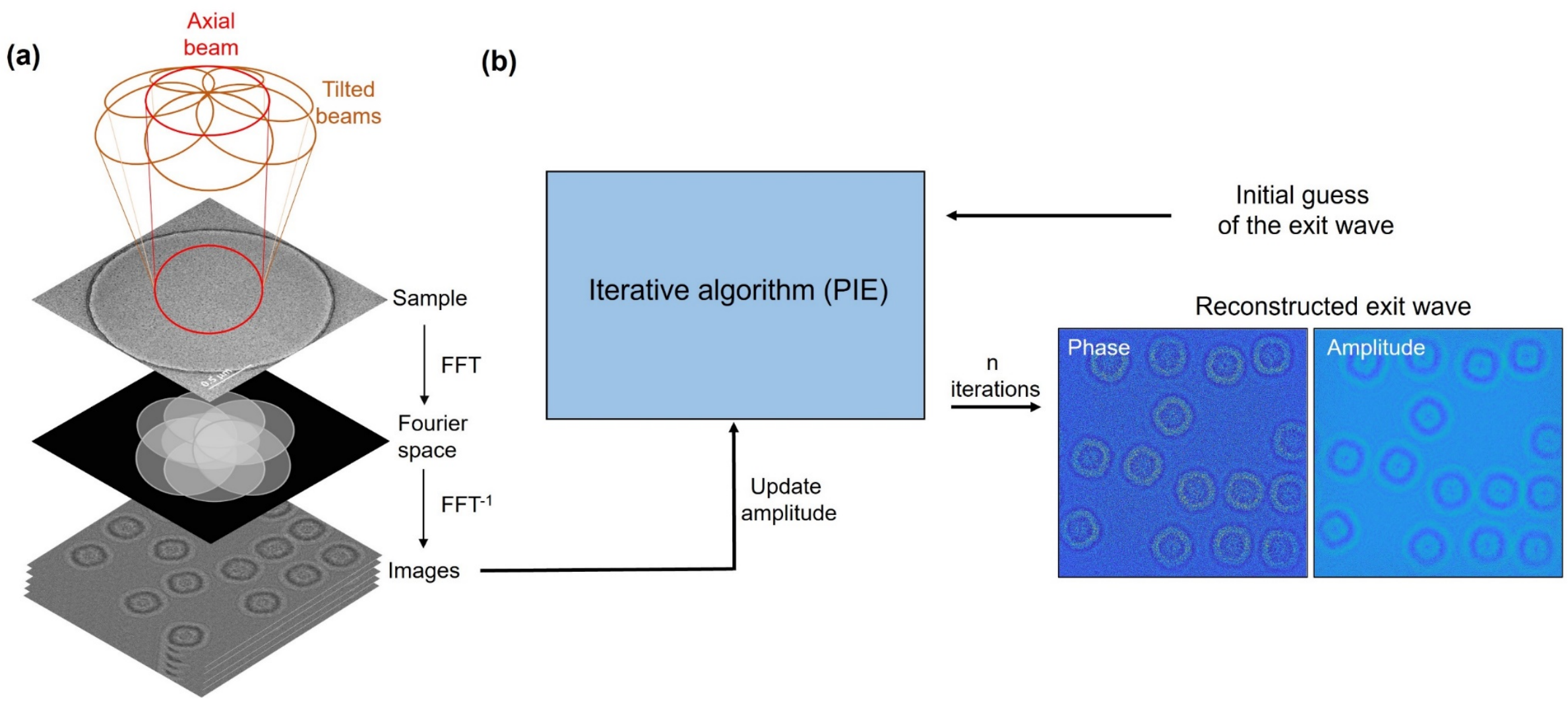

**Fig. 1. Schematic data collection and exit wave reconstruction for electron Fourier ptychography. (a)** Schematic diagram showing data collection. **(b)** Fourier ptychographic exit wave reconstruction using the PIE algorithm. Right bottom corner shows an example of the reconstructed phase and amplitude from simulated apoferritin data.

Image data was recorded at several illumination tilt azimuths with a constant tilt magnitude. For radiation robust samples, the eFP dataset included one axial illumination image and six tilted illumination images. However, for radiation sensitive samples, a reduced dataset containing only four tilted illumination images without an axial image was used as discussed subsequently. Prior to exit wave reconstruction, three additional processing steps were implemented: firstly, the axial defocus was estimated from the cross-correlation[49,50] of the amplitude spectrum of the experimental image and contrast transfer functions (CTFs) calculated for varying defocus values ($C_1$); secondly, images were registered using phase correlation[51]; and thirdly, the tilt-induced shift due to uncorrected residual axial aberrations was compensated. This latter compensation is required as beam tilt introduces an image shift (due to the presence of axial aberrations) in addition to any sample/stage drift during data acquisition. However, these two sources of image shift cannot be separated and hence images need to first be drift-corrected and then compensated for the image shift from the measured aberrations and beam tilt (Supplementary **Text S1**). During the reconstruction of the exit wave, these pre-processed images were used to update the amplitude of the corresponding image wave for each tilted illumination within the PIE algorithm[43]. A flowchart illustrating the PIE algorithm as used for eFP reconstruction is shown in **Fig. 1** and Supplementary **Fig. S1**, while the steps carried out within each PIE iteration are described in Supplementary **Text S2**. It should be noted that in the implementation of the PIE algorithm described here, the amplitude update of the image wave occurs in real space, while that of the exit wave takes place in Fourier space, corresponding to a reversal of the update spaces used in the conventional PIE algorithm.

## Reconstruction from radiation-resistant samples

Simulated datasets of gold particles were initially used to investigate the effects of varying tilt magnitudes on eFP reconstruction. All data were simulated including the effects of the Gatan K2 Summit detector, using the parameters given in **Table 1**. Further details of the process used to simulate the final image intensity are provided in the Methods section. **Figure 2** shows a typical set of reconstructions from simulated data at a fluence of $4.6\times10^5$ $e^-/nm^2$, which is approximately the experimental fluence for the experimental data in **Fig. 3**. A complete eFP dataset of simulated images with a tilt magnitude of 10.0 mrad is shown in Supplementary **Fig. S2**, demonstrating that high resolution information is present in the tilted illumination images but not in the axial illumination image. This is consistent with the beam tilt introducing a directional shift in the effective wave transfer function (**Fig. 2**, **(a)** and **(d)**) and transferring higher resolution information in the tilted illumination images in one direction. The corresponding reconstructions are shown in **Fig. 2**, **(b)-(c)**.

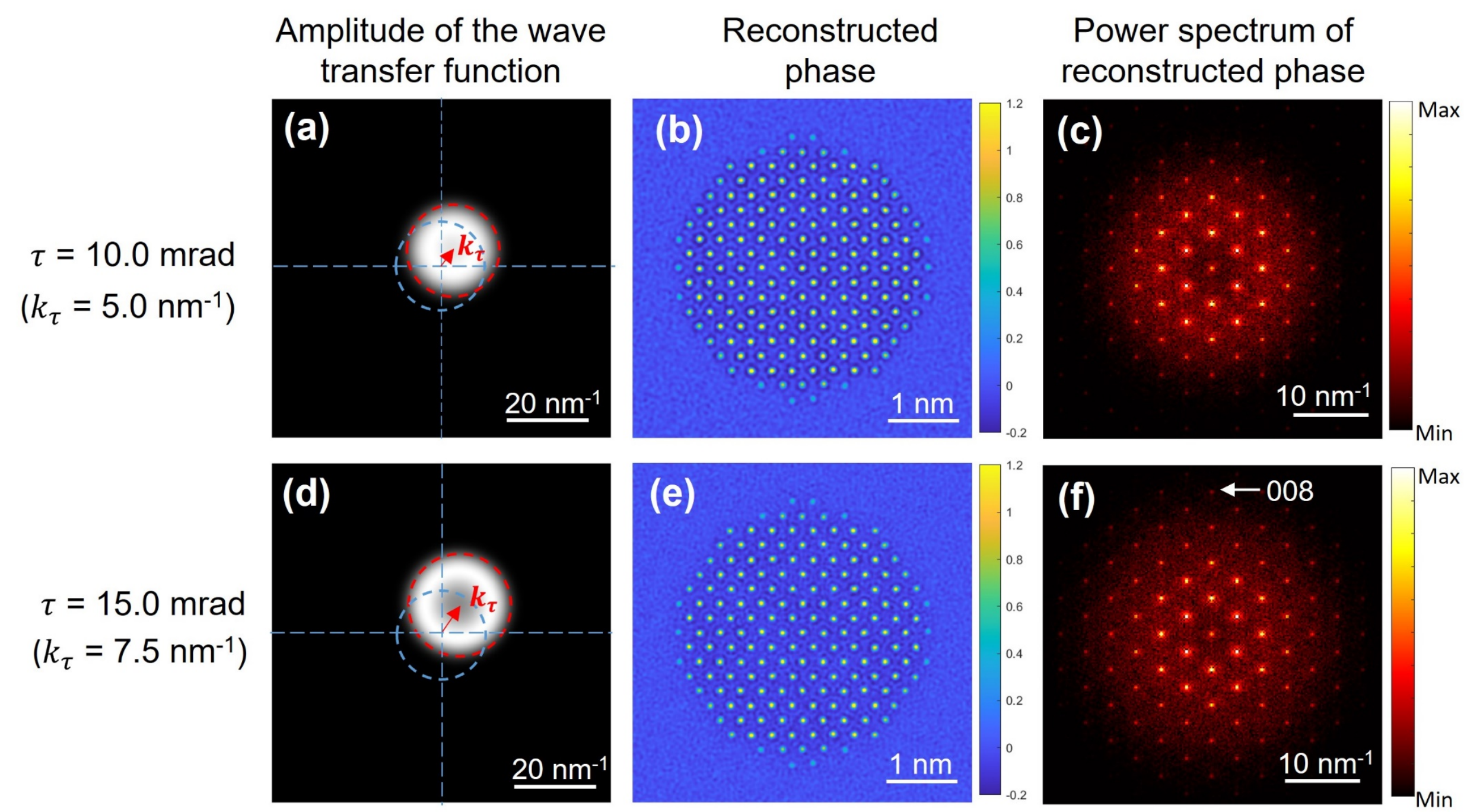

**Fig. 2. Reconstructions of simulated gold particle data with different tilt magnitudes. (a)-(c)** and **(d)-(f)**, Amplitude of the effective wave transfer function, reconstructed phase, and power spectra of the reconstructed phase from datasets with tilt magnitudes of 10.0 mrad and 15 mrad, respectively. These tilt angles give frequency shifts in reciprocal space of 5.0 $nm^{-1}$ and 7.5 $nm^{-1}$, respectively. The gold particle was oriented along <110>. The phase scale bars in **(b)** and **(e)** are in radians. Power spectra intensities are weighted as a power of 0.2 for visualisation of high frequency information.

To extend the resolution in the reconstructed phase, the tilt magnitude can, in principle, be increased. As an example (**Fig. 2**, **(d)-(f)**), an increase in resolution is observed in the power spectrum of the reconstructed phase when the tilt angle is increased to 15.0 mrad with the (008) reflection is present in **Fig. 2(f)** but not in **Fig. 2(c)**. However, when the tilt magnitude increases beyond a critical value, there is a reduction in information transfer at the centre of the wave transfer function. This is caused by partial temporal coherence[23,47], and gives rise to annular information transfer (**Fig. 2(d)**). For the partial temporal coherence function (Supplementary **Text S1**) used in the simulations described here, beam tilt magnitudes of 10 mrad and 15 mrad result in maximum information transfer losses of ~ 10% and ~ 45%, respectively, at the centre of the wave transfer function. This reduction in the wave transfer function could in principle be compensated by modifying the data collection approach, either by increasing the number of tilts at the same magnitude, including data at different defoci at each tilt magnitude or, more effectively, by including additional tilt magnitudes. However, these approaches come at the price of increased complexity in data collection and reconstruction and importantly require increased overall fluence.

**Table 1. Parameters for eFP data simulations.** The notation of the aberration parameters follows that given by Saxton[52]. Positive $C_1$ values are defined as overfocus.

| | Gold particle | Apoferritin | Cry11Aa |
|---|---|---|---|
| Simulation box size (nm$^3$) | 13.5 × 13.5 × 0.8 | 102.4 × 102.4 × 30.0 | 153.6 × 153.6 × 30 |
| Slice thickness (nm) | 0.14 | 1.00 | 1.00 |
| Number of slices | 6 | 31 | 31 |
| Accelerating voltage (kV) | 300 | 300 | 300 |
| Illumination semiangle (mrad) | 0.02 | 0.02 | 0.02 |
| Focal spread (nm) | 4.3 | 8.5 | 8.5 |
| $C_1$ (nm) | 2.5 | -2000 | -2172 |
| $C_3$ (mm) | $-1.9\times10^{-5}$ | 2.7 | 2.7 |
| Pixel size (nm/pixel) | 0.013 | 0.05 | 0.25 |
| Tilt magnitudes (mrad) | 10 and 15 | 5 | 1.9 |
| Total fluence (e$^-$/nm$^2$) | $4.6\times10^5$ | $1\times10^3$, $3\times10^3$, $9\times10^3$, and infinite fluence | $4.5\times10^2$ |
| Number of tilts | 7 | 4, 5, 9, and 13 | 4 |

**Table 2. Parameters for experimental eFP data collection.**

| Specimen | Gold particles | Cry11Aa | Rotavirus |
|---|---|---|---|
| Instrument | JEM-ARM300F2 | JEM-Z300FSC | JEM-Z300FSC |
| Temperature | Ambient | Liquid $N_2$ | Liquid $N_2$ |
| Accelerating voltage (kV) | 300 | 300 | 300 |
| Illumination semi-angle (mrad) | 0.02 | 0.02 | 0.02 |
| Focal spread (nm) | 3.2 | 8.5 | 8.5 |
| $C_1$ (nm) | 5 (11.5)* | -2000 (-2172) | -2000 (-2025) |
| $C_3$ (mm) | 0.0015 | 2.7 | 2.7 |
| Field of view (nm$^2$) | 51.9×53.7 | 1038.8×1074.6 | 400.8×414.8 |
| Pixel size (nm/pixel) | 0.014 | 0.26 | 0.11 |
| Detector | K2 Summit | K2 Summit | K2 Summit |
| Total fluence (e$^-$/nm$^2$) | $4.3\times10^5$ | $4.5\times10^2$ | $2.6\times10^3$ |
| Tilt magnitudes (mrad) | 12.3 | 1.9 | 1.9 |
| Number of tilts | 6 | 4 | 4 |
| Axial illumination (yes/no) | yes | no | no |

*The values in parentheses are the defocus estimated and used in the reconstruction as described in the Methods section.

We applied this method to acquire an experimental dataset from a gold particle using a JEM-ARM 300F2 equipped with two triple hexapole correctors[53]. The gold particle was aligned close to a <110> direction. Parameters for the data collection, processing, and exit wave reconstruction are given in **Table 2** and the Methods section. In the power spectrum calculated from the axial illumination image amplitude (**Fig. 3**, **(a)** and **(b)**), the highest recorded reflection has a resolution of 0.052 nm (I/$\sigma$ ~ 1.3) which corresponds to the {008} reflections. This I/$\sigma$ metric is based on the average intensity of the reflection after background subtraction and the standard deviation of the nearby background. The axial information limit of the microscope (JEM-ARM300F2) used to collect this data is approximately 0.081 nm, based on the measured aberration coefficients and partial coherence limits, corresponding to the point where the information transfer falls below 10%. Although reflections beyond 0.081 nm (e.g., those highlighted by arrows in **Fig. 3(b)**) are present, albeit with reduced low information transfer (< 10%), likely due to dynamical scattering and non-linear interference between the diffracted beams in the strongly scattering gold particle.

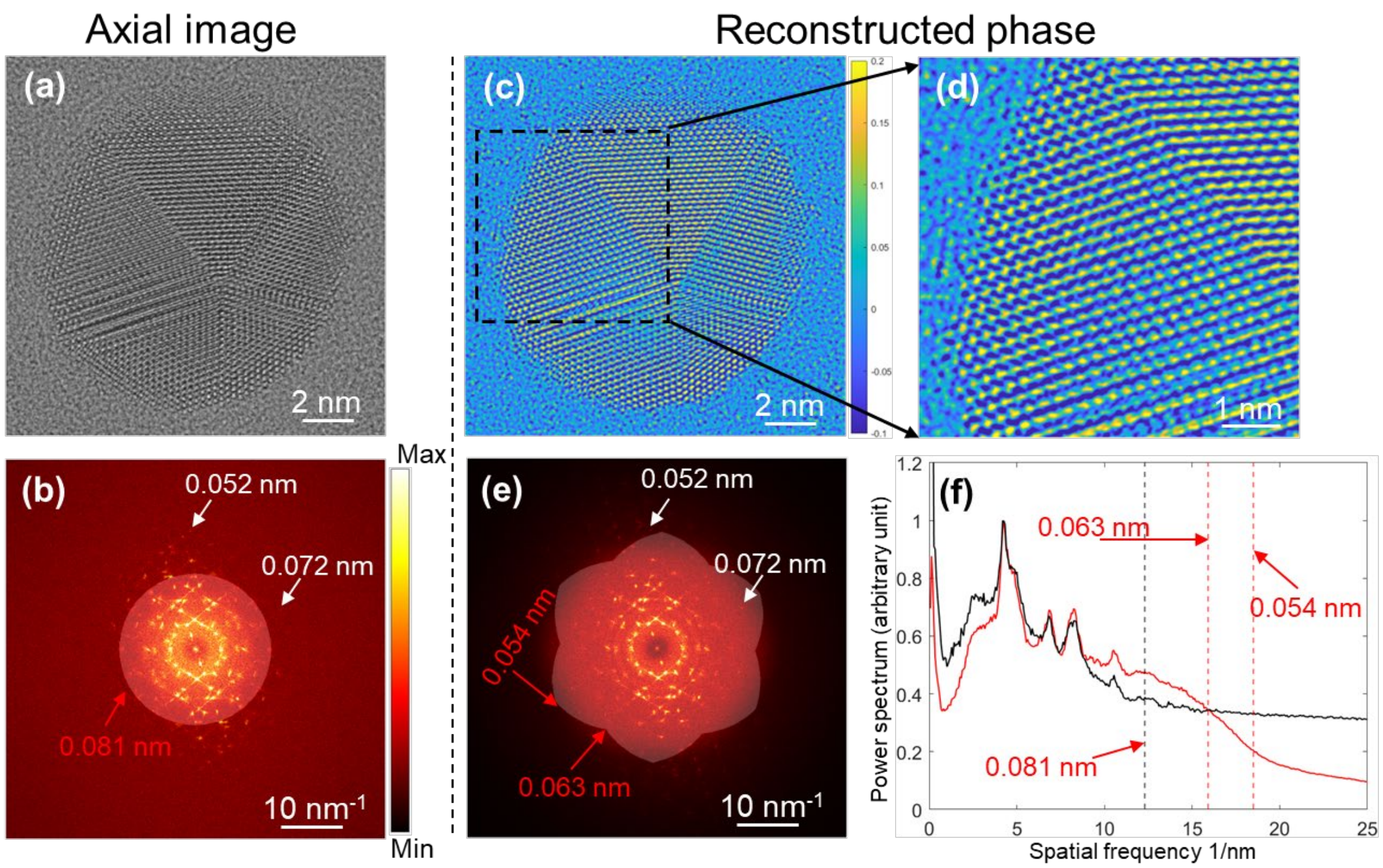

**Fig. 3. Phase reconstruction of gold nanoparticles. (a)-(b)** Axial TEM image and corresponding power spectrum calculated from the image amplitude. The total fluence is the same as that used for the eFP dataset. In **(b)**, an aperture defined by the axial resolution limit (0.081 nm, where information transfer falls below 10%) is overlaid. **(c)-(e)** Phase of the reconstructed exit wave and its power spectrum. **(d)** Enlarged region of **(c)**. In **(e)**, the power spectrum is overlaid with a synthetic aperture using the same 10% information transfer limit and including only spatial frequencies with ≥2 independent measurements. **(f)** Circularly averaged power spectra from **(b)** (black) and **(e)** (red), respectively. Vertical dashed lines indicate the axial resolution limit in **(b)** and the synthetic aperture limit in **(e)**. The phase scale bar in **(c)** is in radians. The intensity of all power spectra is weighted as a power of 0.2 for visualisation of high frequency information and normalized to the strongest reflection at 0.210 nm. For display, phase images in **(c)** and **(d)** are gaussian filtered with standard deviation of 4. The tilt magnitude was 12.3 mrad.

The reconstructed phase and corresponding power spectrum of the gold nanoparticle are shown in **Fig. 3(c)-(f)**. An advantage of Fourier ptychography is the ability to reconstruct the exit wave within a larger effective Fourier space aperture (synthetic aperture) by combining several shifted smaller apertures[32-39]. In this study, the synthetic aperture is defined by the axial aperture (the axial wave transfer function) and six shifted apertures (the effective wave transfer functions under tilted illumination) as illustrated in **Fig. S3**. The complex wavefunction can only be solved at spatial frequencies with at least two independent measurements; therefore, a synthetic aperture containing only those frequencies is most useful (**Fig. 3(e)**). This aperture indicates that the information limit extends to 0.063 nm in all directions and to 0.054 nm midway between neighbouring tilt directions.

The improvement in information transfer is further illustrated by the circularly averaged power spectra of axial image and reconstructed phase over the frequency range 6.3–15.6 $nm^{-1}$ (**Fig. 3(f)**). At low spatial frequencies (0.5–3.6 $nm^{-1}$), the information transfer in the power spectrum calculated from the reconstructed phase is lower than that in the axial image, likely due to suppression by the sinusoidal form of the phase-contrast transfer function. In contrast, the axial image at low frequencies is dominated by phase contrast but also contains contributions from amplitude contrast with a cosinusoidal transfer function, resulting in higher information transfer. Above 15.9 $nm^{-1}$ (0.063 nm), noise dominates and is suppressed in the reconstruction, producing a drop in the circularly averaged power spectrum of the reconstructed phase. In addition, high resolution reflection spots at 0.072 nm and 0.052 nm are clearly visible in the power spectrum calculated from the reconstructed phase (**Fig. 3(e)**), with improved $I/\sigma$ values of 2.3 and 4.7, respectively, compared to 0.5 and 1.3 in the calculated axial power spectrum. Based on this analysis of the power spectra, we demonstrate that eFP can recover phase information with improved transfer at high frequencies compared with axial illumination. A more comprehensive study of this capability was presented by Haigh and co-workers[42], who analysed both the power spectra and atomic separations; here, we focus on demonstrating exit wave reconstruction using eFP and PIE[43,44].

The experimental reconstructed phase shift range is small, whereas, in theory, a significantly larger value is expected for a gold nanoparticle with a diameter of approximately 10 nm. This discrepancy may arise from the Stobbs factor[54–56]. In addition, the gold particle likely adopts the low energy morphology of a Marks decahedron[57], where the relative thickness variation from the edge to the middle is less than for a simple pentagonal bipyramidal morphology.

## Application to beam-sensitive samples

Conventional electron ptychography has been successfully demonstrated for structural studies of radiation sensitive biological samples in a cryo state[14,17–19]. In this section, we describe the application of eFP as an alternative method applied to this sample type. The majority of cryo-EM instruments used in structural biology are not aberration corrected with a finite positive spherical aberration. Imaging biological samples generally also requires a large defocus to transfer low spatial frequencies for enhanced image contrast. Under these conditions, for eFP a tilt magnitude of 27 mrad, calculated using the optimal coupling[23,47] given by $C_1 = -C_3\boldsymbol{\tau}^2$, provides a symmetric partial spatial coherence envelop. However, this large tilt magnitude results in substantial information transfer loss due to partial temporal coherence (Supplementary **Fig. S4**). To illustrate this, we have calculated the effective wave transfer function and coherence envelope functions for beam tilt magnitudes ranging from 0 to 15 mrad, as shown in Supplementary **Fig. S4**. For increasing beam tilt, the diameter of the partial temporal coherence envelope broadens, and a region with a loss of information transfer emerges at the centre of the wave transfer function, consistent with the behaviour described by Supplementary Eq. (3). Increasing the beam tilt also introduces an asymmetry arising from the form of the envelope due to partial spatial coherence which is determined by the gradient of the effective wave aberration function (Supplementary Eq. (4)[23]) Therefore to achieve more uniform information transfer in the effective wave transfer function for tilted illumination, reducing the focal spread, illumination semi-angle, and aberration coefficients are all beneficial. For calculations matching the experimental data collected on a JEM-Z300FSC cryo-microscope (**Table 2**) a small beam tilt is required to minimise the effects of partial coherence. Overall, we have found that a tilt magnitude ≤ 5 mrad offers a practical compromise, resulting in less than 10% information loss and no significant asymmetry in the information transfer (Supplementary **Fig. S4**). However, this small tilt magnitude still provides a resolution improvement over the axial imaging limit. For example, at 300 kV, a beam tilt magnitude of 5 mrad can theoretically extend the resolution of the data for the microscope used from 0.20 nm to 0.13 nm.

In addition, using multiple tilted illumination images with a constant tilt magnitude and varying azimuths can, in theory, increase both the signal to noise ratio for spatial frequencies in the overlap regions and the reconstruction stability due to enhanced overlap. However, for extremely radiation sensitive materials, data must be acquired under low fluence conditions (typically $3\times10^3$ - $4\times10^3$ e/nm$^2$)[58]. This suggests that the number of tilted illumination images that can be included in the reconstruction may also be limited by the overall electron fluence budget. For this reason, we focus on the impact of the total electron fluence, as well as the number of tilted illumination images included in the dataset for reconstruction using eFP. Details of these simulations and reconstructions are given in **Table 1** and in the Methods section. For this analysis, the quality of the reconstructed phase was evaluated using the peak signal-to-noise ratio (PSNR), calculated using a simulated phase at infinite fluence as the ground truth (Supplementary **Fig. S5**). As expected, at infinite fluence, the reconstructed phase from datasets with different numbers of tilts gives equivalent PSNR values as shown in **Fig. 4**, **(a1)** -(**d1)**. At low fluence, adding additional tilted illumination images to the reconstruction lowers the PSNR values of the reconstructed phase (**Fig. 4**). This is attributed to the fact that, for a given total fluence budget, an increased number of tilts results in a reduction in the fluence per image, leading to increased noise in the reconstructed phase.

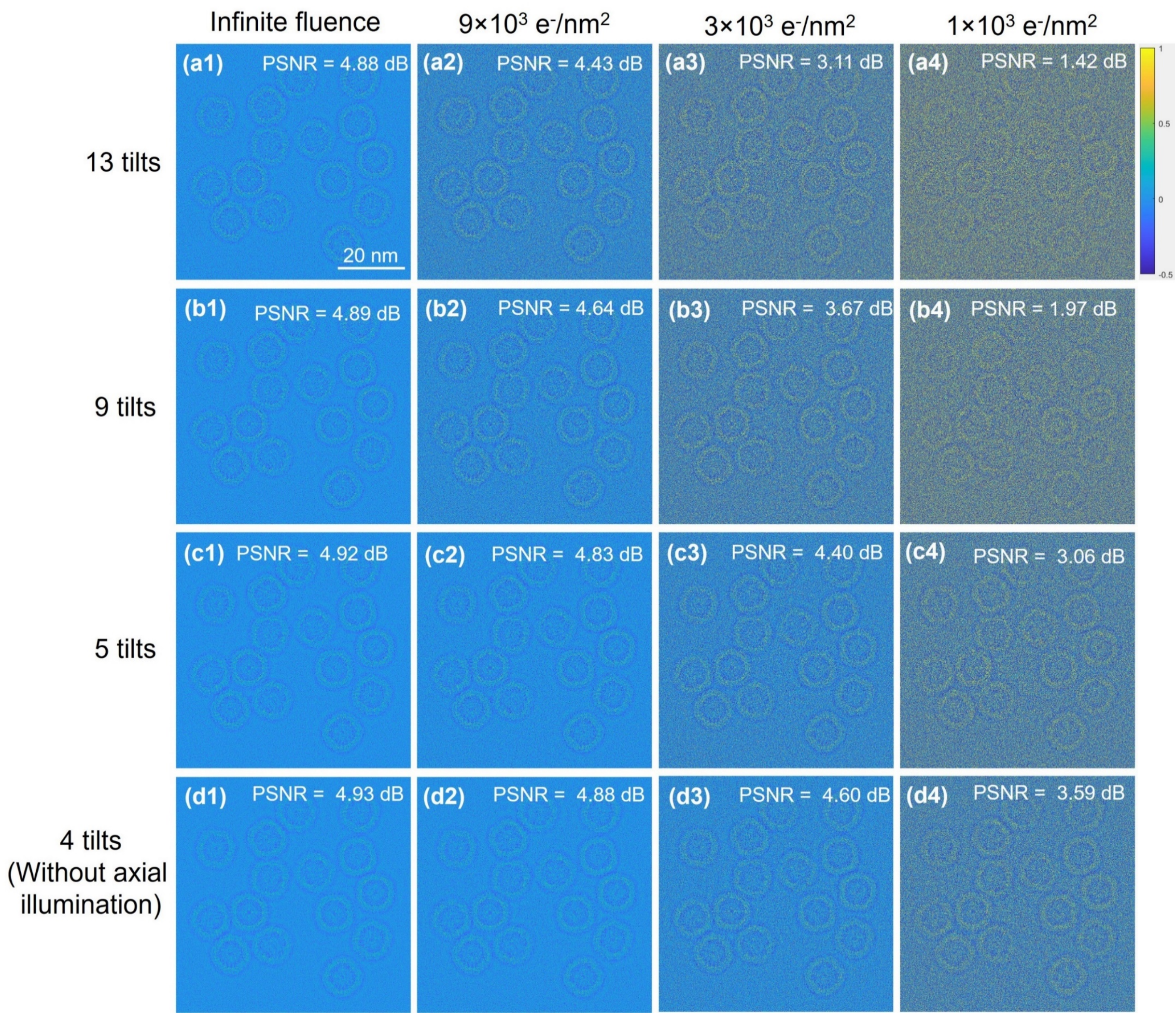

**Fig. 4. Exit wave reconstruction and PSNR of simulated apoferritin data with different numbers of tilted illumination images and varying fluence. (a1)-(a4)**, **(b1)-(b4)**, **(c1)-(c4)** and **(d1)-(d4)**, reconstructed phase and calculated PSNR values from datasets with different total fluence and total number of tilted images. All images use the same scale bar as **(a1)**. The phase scale bar is in radians. Odd numbers for the total tilts include one axial illumination image.

However, insufficient tilts result in azimuthally asymmetric information transfer. The defocus offset between the axial illumination and tilted illumination images is approximately 6.8%, for the tilt magnitude and aberrations used here. Hence, for a low total fluence budget, small defocus offset, and a requirement for azimuthally symmetric information transfer, we have opted for collection of low fluence eFP datasets using only four tilted illumination images.

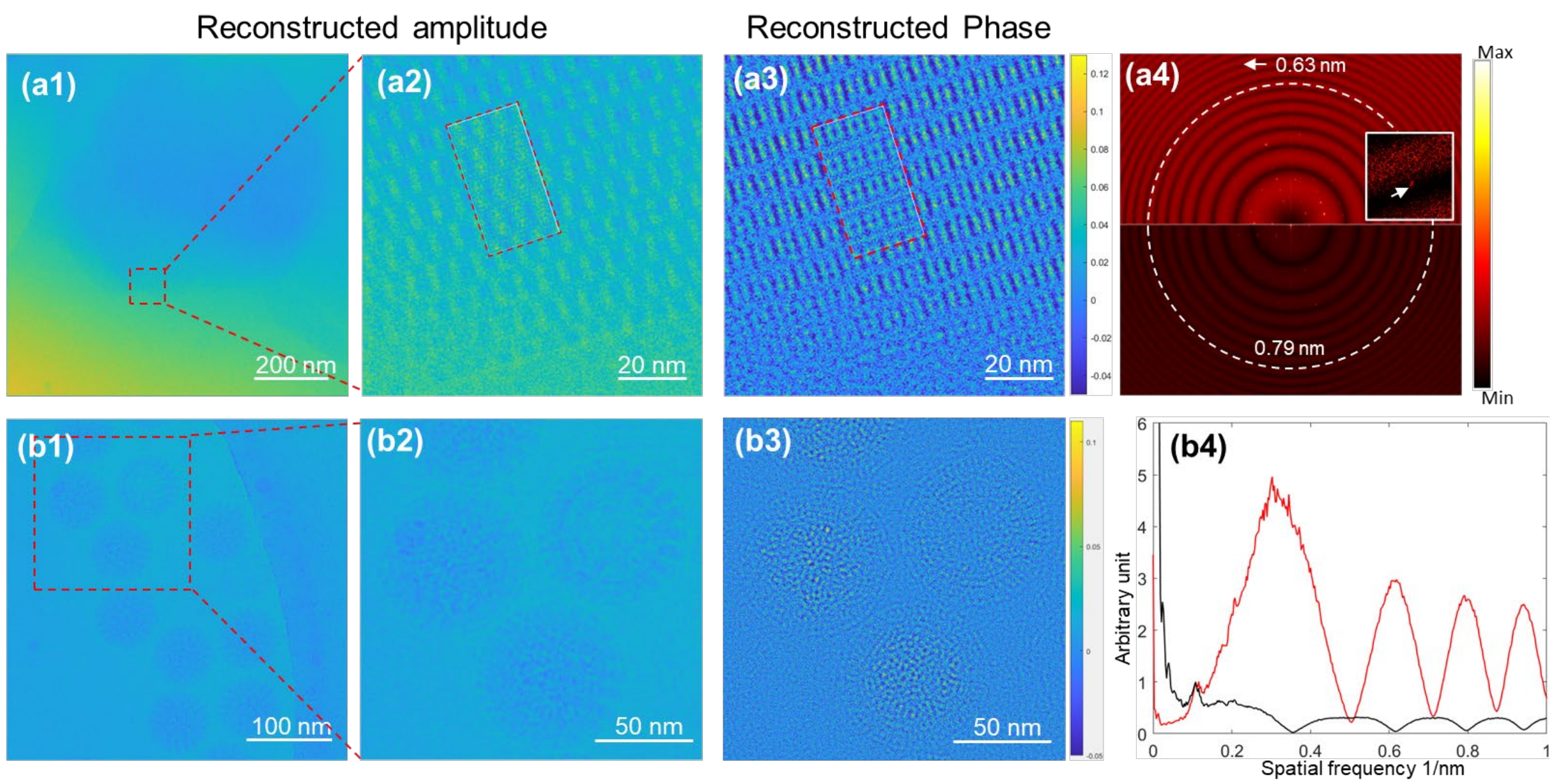

**Fig. 5. Exit wave reconstruction of Cry11Aa and rotavirus samples. (a1)-(a4)** Reconstructed amplitude, phase, and calculated power spectra of the experimental Cry11Aa data. Insets to **(a2)** and **(a3)** shows the reconstructed amplitude and phase from simulated Cry11Aa data using the experimental parameters. The upper section of **(a4)** shows the power spectrum of the reconstructed phase, while the lower section displays the power spectrum of the reconstructed amplitude. Inset to (**a4**) is an enlarged view of the reflection at the resolution of 0.63 nm in the power spectrum of reconstructed phase. **(b1)-(b3)** Reconstructed amplitude and phase of rotavirus particles. **(b4)** Plots of circular averages of the amplitude of the Fourier transform of the reconstructed phase (red) and amplitude (black), respectively. Plots are normalised with respect to the intensities of peaks at 0.12 $nm^{-1}$, corresponding to the separation between the capsid trimers of viral protein 6 (VP6)[17]. The phase scale bars in images **(a3)** and **(b3)** are in radians. The reconstructed amplitude and phase in **(a1)-(a3)** and **(b1)-(b3)** are filtered using Gaussian filters with standard deviations of 2 and 3, respectively.

The potential of eFP for studies of radiation-sensitive biological samples was further investigated using vitrified specimens of insecticidal protein Cry11Aa crystals and rotavirus particles. All eFP datasets were collected using an uncorrected cryo-microscope (JEM-Z300FSC) operated at 300 kV, with $C_3$ = 2.7 mm. Further details of the data collection parameters are given in **Table 2** and the reconstructions are described in the Methods section. For the dataset used for the reconstruction shown in **Fig. 5**, **(a1)-(a4)**, the Cry11Aa crystal was approximately orientated along [010]. Lattice fringes are observed in both the amplitude and phase of the reconstructed exit wave (**Fig. 5**, **(a1)-(a3)**), while the reconstructed phase also shows structural detail at higher resolution (0.63 nm) which is absent in the amplitude for the same I/$\sigma$ threshold (>2) ((**Fig. 5(a4)**)). Moreover, the experimental reconstructed phase and amplitude (**Fig. 5, (a2)** and **(a3)**) both qualitatively agree with simulated data for equivalent imaging conditions with details given in **Table 1** and in the Methods section. The reconstruction from the rotavirus data shows that low frequency information (< 0.12 $nm^{-1}$) is mainly transferred in the amplitude, whereas higher resolution structural details are primarily captured in the phase (**Fig. 5**, **(b1)-(b4))** as expected given the form of the respective transfer functions.

For imaging of weakly scattering specimens with thicknesses on the order of a few hundred nm or less, the image contrast is dominated by phase shifts, while the amplitude contrast is weak. However, the contrast transfer function modulates both phase and amplitude information through sine and cosine terms, respectively (as shown in **Fig. 5(a4)** and **(b4)**) and consequently amplitude information is transferred more effectively at low spatial frequencies. However, the amplitude information decreases at higher frequencies and is not transferred. In contrast, phase information is suppressed near zero spatial frequency but is more strongly transferred at higher frequencies. We note that oscillations in the transfer function are clearly visible in the power spectra of the reconstructed phase and amplitude in **Fig. 5(a4)** and **(b4)**. This is primarily due to a high defocus (~ 2 μm) which introduces significant phase shifts that are not compensated by those due to a positive spherical aberration in the data in **Fig. 5** and by the limited number of tilted illumination images included in the reconstruction. In contrast the data in **Fig. 3** does not show these effects due to a corrected spherical aberration and values for the defocus and illumination tilt that compensate any residual phase shifts[23].

## Discussion

Reconstruction using tilted illumination can provide resolution beyond the limit for axial illumination and also demonstrates improved low spatial frequency information transfer[23,42]. For conventional ptychography Seki and co-workers have evaluated several STEM coherent imaging modes using a noise normalised phase contrast transfer function[59]. A comparison of eFP, based on real space images to STEM based methods that record diffraction patterns, would be valuable future addition but is beyond the scope of this work.

Due to the tilted illumination acquisition geometry, the synthetic aperture of the reconstructed exit wave also exhibits anisotropic information transfer. Along the illumination tilt directions, high resolution is achieved, but with a data redundancy of one which does not satisfy the oversampling requirements for reconstruction. Between the illumination tilt directions, the resolution is lower, but the data redundancy is greater than one (as illustrated in **Fig. S6**). For frequencies where only a single measurement is available, reconstruction of the complex wavefunction is not possible. Therefore, in practice, only an aperture containing spatial frequencies with data redundancy ≥ 2 are used, as shown in **Fig. 3(e)**. The reflection at 0.072 nm in **Fig. 3(e)** lies within a region with two measurements (**Fig. S3**) and moreover, spatial frequency regions with only a single measurement are small due to the small beam tilt used (**Fig. S6**).

Sample damage ultimately limits the resolution of all imaging and phase retrieval methods when applied to radiation sensitive specimens. For the specific case of eFP, this limits the number of images that can be collected and used in the reconstruction, due to a small finite fluence budget. We also expect that high-resolution information will be better transferred in the early images in a series and will degrade due to radiation damage as the dataset accumulates. This aspect of data acquisition is well known for conventional cryo-EM and can be overcome using a fractionation scheme[60]. Conventional phase contrast image formation for vitrified biological samples is often described within the weak object approximation[61,62] with a small global phase shift added to the phase contrast transfer function to account for the small amplitude contrast[49,63,64]. However, this global phase shift is not easy to experimentally determine precisely, and an empirical value is often used[49,62]. In contrast, ptychographic reconstruction recovers the complex exit wave, eliminating this requirement for an empirical compensation of the phase shift.

Our results show that the low-frequency information, related to the overall object shape, is transferred predominantly in the amplitude. This suggests the possibility of using the reconstructed amplitude information for locating and aligning protein particles or cellular structures and using the reconstructed phase information for high-resolution structural analysis. As eFP is based on imaging using plane-wave illumination and controlled beam tilt, its implementation requires no instrument modifications and hence can be used in all conventional cryo-EM instruments. Finally, eFP can be seamlessly integrated with established data collection and processing software[65,66], making it an accessible and efficient approach for high-resolution phase reconstruction.

# Methods

## Data simulation

<u>Gold particle data</u>

The gold nanoparticle model used for the eFP data simulations was built from a gold crystal with a cubic unit cell; $a = b = c = 0.41$ nm and $\alpha = \beta = \gamma = 90.0°$. The nanoparticle was oriented along a <110> direction. Exit waves at different illumination tilts were simulated using the multislice method[61,67–69]. The slice potential was generated using the MULTEM[70,71] code. Image waves were calculated by applying the wave transfer function to the simulated exit waves, and the final images were calculated from the squared magnitude of the image waves. For the final image intensity simulation, a total fluence of $4.6\times10^5$ $e^-/nm^2$ was used. For all calculations the detective quantum efficiency and the noise transfer function of a K2 Summit detector were included, with parameters taken from previous publications[72–74]. Two eFP datasets were simulated at tilt magnitudes of 10 mrad and 15 mrad, respectively. Each dataset consisted of one axial illumination image and six tilted illumination images with the same tilt magnitude and equally spaced azimuths. Full details of the simulation parameters are given in **Table 1**.

<u>Apoferritin data</u>

Apoferritin datasets were simulated in a box of size 102.4 nm × 102.4 nm × 30.0 nm with the apoferritin particles generated from the structure model PDB ID 7A6A. The data simulation process was similar to that described for the gold particle simulation. One important difference is that the slice potential was generated using Parakeet[75], which is a modification of MULTEM[70,71]. The amorphous ice was represented by Fourier-filtered noise model in Parakeet[76] which has been shown to give equivalent results to a full DFT simulation. The slice thickness was set to 1.0 nm[77]. Full details of the simulation parameters are given in **Table 1**.

<u>Cry11Aa data</u>

Cry11Aa crystal datasets were simulated in a box of size 153.6 nm × 153.6 nm × 30 nm. A Cry11Aa supercell, consisting of 15×1×8 unit cells, was generated using the structure from PDB ID 7QX4. The supercell was oriented along [010], as shown in Supplementary **Fig. S7**, aligned with the z-axis of the simulation box. The data simulation process is identical to that described for the apoferritin simulation. Full details of the simulation parameters are given in **Table 1**.

## Sample preparation

<u>Gold particle specimen</u>

A standard sample grid coated with a thin amorphous germanium film (~ 2-10 nm) and gold nanoparticles (~ 5-20 nm in diameter) was used for the experiments with results shown in **Fig. 3**.

Rotavirus specimen
Rotavirus samples were prepared following a previously described method[17], with only a brief summary provided here. A suspension of rotavirus double layered particles was diluted in 20 mM Tris HCl, 1 mM EGTA to 8 mg/ml, and 4 µl of the rotavirus suspension was applied to a plasma-cleaned Quantifoil holey carbon grid (300 mesh Cu R2/2). The grid was then blotted for 5s and plunged into liquid ethane using a Gatan CP3 semi-manual plunger at an ambient humidity of 80%. The plunge-frozen grid was then transferred and stored in liquid nitrogen before loading into the microscope.

Cry11Aa protein crystal specimen
Production and purification of Cry11Aa nanocrystals were performed as previously described[78]. Purified crystals were stored in ultrapure water at 4 °C until use. Samples were prepared from 100 µl of Cry11Aa crystal suspensions diluted in 1 ml of $dH_2O$ before being vortexed and then further diluted 10-fold with a solution of 10% glycerol. Vitrified grids were prepared by applying 3 µl of the crystal suspension to the carbon side of a freshly glow-discharged (60s 20 mA) Quantifoil grid (300 mesh Cu R2/2). Excess solution was removed using a Vitrobot mark IV at room temperature with the humidifier off. Parameters used for blotting were 30 s waiting time, 20 s blotting with a blotting force of 20, and 1 s draining before plunge freezing.

## Experimental data collection

The gold nanoparticle dataset was collected on a JEM-ARM 300F2 microscope equipped with two triple hexapole aberration correctors[53] and operated at 300 kV. The dataset was collected using plane-wave illumination with axial aberrations corrected to third order in the wave aberration function (Supplementary **Table S1**). A simple DM-python script was used to control the data collection. The beam tilt was calibrated in diffraction mode with further details given in Supplementary **Text S3** and **Table S2**. The experimental dataset consisted of one axial image and six tilted illumination images, with a tilt magnitude of 12.3 mrad. The exposure time for each tilt illumination was 2s, and the beam was blanked between illumination changes. Images were recorded on a K2 Summit camera in counting mode with a binning factor of 1.

Datasets from a Cry11Aa crystal and rotavirus particles were collected on a JEM-Z300FSC cryo-microscope at 300 kV. Images were recorded on a K2 Summit camera in dose fractionation mode (separating a single image exposure into multiple subframes). An in-column Omega filter was used to zero loss filter the images with a slit width of 16 eV. Tilt calibration was carried out as described in Supplementary **Text S3** with calibration values given in Supplementary **Table S2**. These datasets only included four tilted illumination images without an axial illumination image. The exposure time for each illumination was 2 s for Cry11Aa data and 1.3 s for rotavirus data and the beam was blanked between illumination changes.

Full details of the data collection parameters for all datasets are given in **Table 2**. The beam tilt modifies the axial aberration coefficients, as discussed in Supplementary **Text S1** and fully described in previous studies[47,52]. To minimise this effect, all axial aberrations were electron optically corrected to third order for data collected on the JEM-ARM 300F2 microscope while off axial aberrations were not considered as the field of view for imaging was small (< 1 µm × 1 µm). For data collected on JEM-Z300FSC cryo-microscope a simple coma-free alignment was performed following two-fold astigmatism correction.

## Ptychographic reconstruction

Prior to reconstruction, defocus estimation, image registration and tilt-induced shift compensation were carried out as already described. For the gold particle data, the defocus estimation was performed using cross-correlation[49,50] between the amplitude spectrum of an experimental image and a calculated CTF with varying $C_1$. The defocus of the gold particle data was estimated to be 11.5 ± 0.25 nm. For the remaining data, MotionCor2[79] was used for motion correction within each tilted illumination image stack and CTFFIND4[50] was used for defocus estimation. The axial defocus of the Cry11Aa data was estimated from an axial illumination image recorded from an amorphous region near the target crystal. The axial defocus of the rotavirus data was estimated from the average effective defocus of the four tilted illumination images and the theoretically calculated defocus change introduced by the beam tilt as described in Supplementary **Text S1**. The average defocus and standard error were calculated from the estimated defocus values of 4×4 patches from each image above that were used as input to CTFFIND4. Outliers exceeding three times the median absolute deviation were excluded. The estimated axial defocus values were -2172 ± 19 nm for the Cry11Aa dataset and -2025 ± 4 nm for the rotavirus dataset. Images at different tilt illuminations were finally registered using phase correlation[51] and compensated for the calculated image shift induced by uncorrected aberrations (with $C_1$ and $C_3$ only considered) and beam tilt.

Exit wave reconstruction was performed using the PIE algorithm[43], as summarised in Supplementary **Text S2**. The initial estimate of the exit wave was generated in real space with unity amplitude and zero phase. During the exit wave update, a step decay scheme was used with a decay rate of 0.5 for every 10 iterations. The initial step size for updates was set to 0.1 for both simulated and experimental data and the final reconstruction converged after 50 iterations. For exit wave reconstruction of the experimental datasets from the biological samples, an upsampling scheme was applied as described in Supplementary **Text S4**.

## Data availability

All data needed to evaluate the conclusions in the paper are available in the main text or the supplementary materials. The data and codes used for the reconstructions have been deposited in the Zenodo database and can be downloaded at: https://doi.org/10.5281/zenodo.11482815.

## Acknowledgments

We thank T. Starborg, J. Barnard, and M. Yusuf from the Rosalind Franklin Institute for technical assistance. We also thank K. Treder from the University of Oxford for his support in gold particle model building. We acknowledge P. Wang from University of Warwick, E. Liberti from the Rosalind Franklin Institute, and G. Zheng and P. Song from University of Connecticut for relevant discussions.

## Funding

This work was funded by the UK Research and Innovation, Engineering and Physical Sciences Research Council. This research was also supported by the Agence Nationale de la Recherche (grants ANR-17-CE11-0018-01 and ANR-22-CE11-0016-01 to J.-P.C). IBS[5] acknowledges integration into the Interdisciplinary Research Institute of Grenoble (IRIG, CEA).

## Author contributions

A.I.K and C.H conceived the overall project. J.Z devised the reconstruction codes, designed and performed the experiments, and carried out the reconstructions and analyses. Al.M and Am.M provided support in code development. J.M.P and I.L provided support in data simulations. M.G-J and J.S.K provided support in results analysis and discussions. M.B, D.S, E.A and J.-P.C prepared the biological samples. J.Z, A.I.K, and C.H prepared the manuscript with contributions from all authors.

## Competing interests
The authors declare no competing interests.

# Supplementary Material for

## Electron Fourier ptychography for phase reconstruction

Jingjing Zhao *et al.*

Corresponding author: Jingjing Zhao, jingjing.zhao@rfi.ac.uk

**This file includes:**

Supplementary Text S1 to S4
Figs. S1 to S10
Tables S1 to S2
References (1 to 9)

## Supplementary Text

Text S1. Effective wave transfer function

The effective wave transfer function $w'(\boldsymbol{k},\boldsymbol{k_\tau})$ under tilted illumination is described by[1,2]:

$$w'(\boldsymbol{k},\boldsymbol{k_\tau}) = E_t'(\boldsymbol{k},\boldsymbol{k_\tau})E_s'(\boldsymbol{k},\boldsymbol{k_\tau})\exp(-i\chi'(\boldsymbol{k},\boldsymbol{k_\tau})) \quad (1)$$

where $\chi'(\boldsymbol{k},\boldsymbol{k_\tau})$ is the effective aberration function, and $E_t'(\boldsymbol{k},\boldsymbol{k_\tau})$ and $E_s'(\boldsymbol{k},\boldsymbol{k_\tau})$ are the partial temporal and spatial coherence envelope functions. In Eq.(1), $\boldsymbol{k}$ is a two-dimensional vector in reciprocal space, and $\boldsymbol{k_\tau}$ is the two-dimensional wave vector of the tilted incident beam.

These three functions are given by:

$$\chi'(\boldsymbol{k},\boldsymbol{k_\tau}) = \chi(\boldsymbol{k}+\boldsymbol{k_\tau}) - \chi(\boldsymbol{k_\tau}) \quad (2)$$

$$E_t'(\boldsymbol{k},\boldsymbol{k_\tau}) = exp\left\{-(\tfrac{\pi\Delta\lambda}{2})^2\left[(\boldsymbol{k}+\boldsymbol{k_\tau})^2 - \boldsymbol{k_\tau}^2\right]^2\right\} \quad (3)$$

$$E_s'(\boldsymbol{k},\boldsymbol{k_\tau}) = exp\left\{-(\tfrac{\theta}{2\lambda})^2|\nabla\chi'(\boldsymbol{k},\boldsymbol{k_\tau})|^2\right\} \quad (4)$$

where $\Delta$ is the $e^{-1}$ half-width value of the focal spread distribution, $\theta$ is the illumination semiangle and $\lambda$ is the electron wavelength. Details of the effective aberration coefficients have been previously described[1–3]. Here we define an effective image shift $A_0'$ and effective defocus $C_1'$ as used in the main text as:

$$A_0' = A_0 + A_1\tau^* + C_1\tau + A_2\tau^{*2} + \tfrac{1}{3}B_2^*\tau^2 + \tfrac{2}{3}B_2\tau^*\tau + C_3\tau^*\tau^2 \quad (5)$$

$$C_1' = C_1 + \mathrm{Re}(\tfrac{4}{3}B_2\tau^*) + 2C_3\tau^*\tau \quad (6)$$

where $A_0$, $A_1$, $C_1$, $A_2$, $B_2$, and $C_3$ are axial aberration coefficients corresponding to image shift, two-fold astigmatism, defocus, three-fold astigmatism, axial coma, and spherical aberration, respectively.

This notation of the aberration coefficients follows the convention given by Saxton[3] and we write the beam tilt $\tau$ in complex form as $\tau = t_x + it_y$.

Text S2. Iterative reconstruction

The update steps for one iteration of the modified Ptychographic Iterative Engine (PIE)[4,5] algorithm used in this work are similar to those in conventional PIE-base ptychographic reconstruction. The key difference is that, in Fourier ptychography, the amplitude update of the image wave occurs in real space, while the exit wave update takes place in Fourier space. In contrast, conventional ptychographic reconstruction updates the image wave in Fourier space and the exit wave in real space. Hence, the basic steps for one iteration can be summarised as in **Fig. S1** and described as follows:

1. An initial estimate of the exit wave $\psi_{ex}(\boldsymbol{r})$ is generated, where $\boldsymbol{r}$ is a two-dimensional vector in real space. An exit wave $\psi_{ex}(\boldsymbol{r})$ with unity amplitude and zero phase was used as the initial estimate. We further assumed that the input images can be represented as $m \times m$ matrices. In this formalism the initial exit wave $\psi_{ex}(\boldsymbol{r})$ needs to be upsampled to $n \times n$ pixels to provide a sufficiently large array to stitch the shifted information transfer domains (arising from the illumination tilt in Fourier space). When the Nyquist sampling criterion is satisfied, the relationship between $m$ and $n$ is given by: $n > m + |\boldsymbol{k_\tau}|(m * p)$ where $p$ is the pixel size and $\boldsymbol{k_\tau}$ is the two-dimensional wave vector of the tilted incident illumination direction. For example, if the input image size is $m \times m$ = 1024×1024 pixels, and the amplitude of the beam tilt wave vector is $|\boldsymbol{k_\tau}| = 5\ nm^{-1}$, for a pixel size of 0.1 nm/pixel, the image size of the initial exit wave should be larger than 1536×1536 pixels. For simplicity we have used an initial estimate with two-fold upsampling ($n = 2m$). A numerical image size of 2m × 2m pixels for the exit wave $\psi_{ex}(\boldsymbol{r})$ was chosen to ensure a sufficiently large array for stitching the shifted information transfer domains in Fourier space as shown in Supplementary **Fig. S6(a)**. The effective reconstructed region (synthetic aperture) in Fourier space is defined by these stitched domains, as illustrated in **Fig. S6(b)** and **(c)**. The tilt magnitudes and other calculation parameters used in **Fig. S6(b)** and **(c)** are the same as those used in **Fig. 3** and **Fig.5** in the main text. Based on these calculations, the effective reconstructed pixel numbers in **Fig. S6(b)** and **(c)** are approximately 1.2m$^2$ and 0.7m$^2$, respectively.

When the data does not meet the Nyquist sampling criterion, an upsampling scheme can be included in the reconstruction to suppress unwanted aliasing. The details of this upsampling scheme are described in **Text S3**. In summary, we assume an upsampling ratio $l$, and therefore the upsampled input image size is given as $m' \times m'$ where $m' = m * l$. For this condition, $n > m' + |\boldsymbol{k_\tau}|(m * p))$ needs to be satisfied.

2. The $j^{th}$ exit wave $\Psi_{ex_j}(\boldsymbol{k}, \boldsymbol{k_\tau})$ is obtained from the initial exit wave $\Psi_{ex}(\boldsymbol{k})$ in Fourier space. Here, $\Psi_{ex_j}(\boldsymbol{k}, \boldsymbol{k_\tau})$ is one Fourier domain in $\Psi_{ex}(\boldsymbol{k})$ with its position defined by the wave vector of the tilted incident beam $\boldsymbol{k_\tau}$. The index $j$ indicates the update corresponding to the $j^{th}$ tilted illumination.

3. The $j^{th}$ image wave is generated in real space as: $\psi_{im_j}(\boldsymbol{r}, \boldsymbol{k_\tau}) = \mathcal{F}^{-1}(\Psi_{im_j}(\boldsymbol{k}, \boldsymbol{k_\tau})) = \mathcal{F}^{-1}[\Psi_{ex_j}(\boldsymbol{k}, \boldsymbol{k_\tau})w(\boldsymbol{k})]$ with $w(\boldsymbol{k})$ the wave transfer function.

4. The amplitude of the $j^{th}$ image wave is updated from the recorded $j^{th}$ image $I_j(\boldsymbol{r})$, while keeping the phase unchanged: $\psi'_{im_j}(\boldsymbol{r}, \boldsymbol{k_\tau}) = \sqrt{I_j(\boldsymbol{r})} \cdot \frac{\psi_{im_j}(\boldsymbol{r}, \boldsymbol{k_\tau})}{|\psi_{im_j}(\boldsymbol{r}, \boldsymbol{k_\tau})|}$.

5. The updated $j^{th}$ image wave is calculated in Fourier space as: $\Psi'_{im_j}(\boldsymbol{k}, \boldsymbol{k_\tau}) = \mathcal{F}(\psi'_{im_j}(\boldsymbol{r}, \boldsymbol{k_\tau}))$.

6. The exit wave is updated within the PIE algorithm as:
$$\Psi'_{ex_j}(\boldsymbol{k}, \boldsymbol{k_\tau}) = \Psi_{ex_j}(\boldsymbol{k}, \boldsymbol{k_\tau}) + \alpha \frac{w^*(\boldsymbol{k})}{|w(\boldsymbol{k})|^2_{max}} (\Psi'_{im_j}(\boldsymbol{k}, \boldsymbol{k_\tau}) - \Psi_{im_j}(\boldsymbol{k}, \boldsymbol{k_\tau}))$$
where $\alpha$ is the updating step size and $w^*(\boldsymbol{k})$ is the conjugate of $w(\boldsymbol{k})$. A step decay schedule was applied to assist convergence of the algorithm and to avoid local minima. The decay ratio used in this work was 0.5 for every 10 iterations which provide effective convergence after 50 total iterations.

7. Steps 2–6 are repeated to include all other tilted illumination images in the update of the exit wave.

Text S3. Beam tilt calibration

Calibration of the beam tilt magnitude and orientation is essential for ptychographic data acquisition and reconstruction. Tilt calibration was carried out in diffraction mode with the same illumination convergence used for data collection. The camera length was calibrated using known spacings in diffraction patterns recorded from a polycrystalline gold film. Diffraction datasets were collected with beam tilts applied in four approximately orthogonal directions corresponding to the tilt coil axes (+x, -x, +y, -y), using five tilt magnitudes per direction with a constant tilt step size. The position of the direct beam was used to measure the tilt magnitude and the orientation relationship between the beam tilt and image/detector coordinates. The measured calibrations are given in **Fig. S8** and **Table S2**.

Text S4. Upsampling

For a frequency ($f_{inf}$) that is higher than the Nyquist frequency ($f_{Nyq}$) in the recorded image, aliasing will occur. As for all imaging systems, the field of view (FOV) and resolution is an intrinsic trade off in TEM[6]. This implies that when the magnification is decreased to achieve a large FOV, the image resolution will be limited by the Nyquist sampling frequency (determined by the detector pixel pitch) and further constrained by aliasing. Hence, reconstruction of the exit wave from an eFP dataset that does not meet the Nyquist sampling criterion (**Fig. S9**), will show aliasing artifacts in the recovered exit wave as shown in **Fig. S10**, **(b1)-(b2)** and **(d1)-(d2)**. To overcome this, an upsampling scheme was applied to the update of the exit wave within the PIE algorithm. This is related to an upsampling scheme previously used in X-ray ptychography[7] and optical Fourier ptychography[8]. The key feature of this scheme is the assumption of a large synthetic detector ($m' \times m'$ pixels) with a reduced pixel size that satisfies the Nyquist criterion and the use of the original data ($m \times m$ pixels) to update the amplitude of the upsampled and calculated image wave. In the amplitude update, an amplitude correction matrix $C_{m',m'} = \sqrt{(U_{lanczos3}\left\{\frac{I_{m,m}}{|O_{m,m}|_s^2}\right\})}$ was used, where the $I_{m,m}$ is the recorded image intensity, $\left|O_{m,m}\right|_s^2$ is the calculated and binned intensity from the calculated image wave, and $U_{lanczos3}$ is a lanczcos-3 interpolation. This amplitude correction matrix was applied to the calculated image wave, replacing step 4 (**Text S2**), as $\psi'_{im_j}(\boldsymbol{r}, \boldsymbol{k_\tau}) = \psi_{im_j}(\boldsymbol{r}, \boldsymbol{k_\tau}) \cdot C_{m',m'}$. Using this upsampling scheme, aliasing artifacts are significantly reduced as shown in **Fig. S10**, **(c1)-(c2)** and **(e1)-(e2)**. However, aliasing cannot be eliminated, especially at the high spatial frequencies as highlighted in **Fig. S10**. To mitigate this, a low pass filter with a highest frequency pass of $2/3 f_{Nyq}$ can be applied to the final reconstructed exit wave as is commonly used in the processing of conventional single particle cryo-EM images to avoid interpolation errors[9].

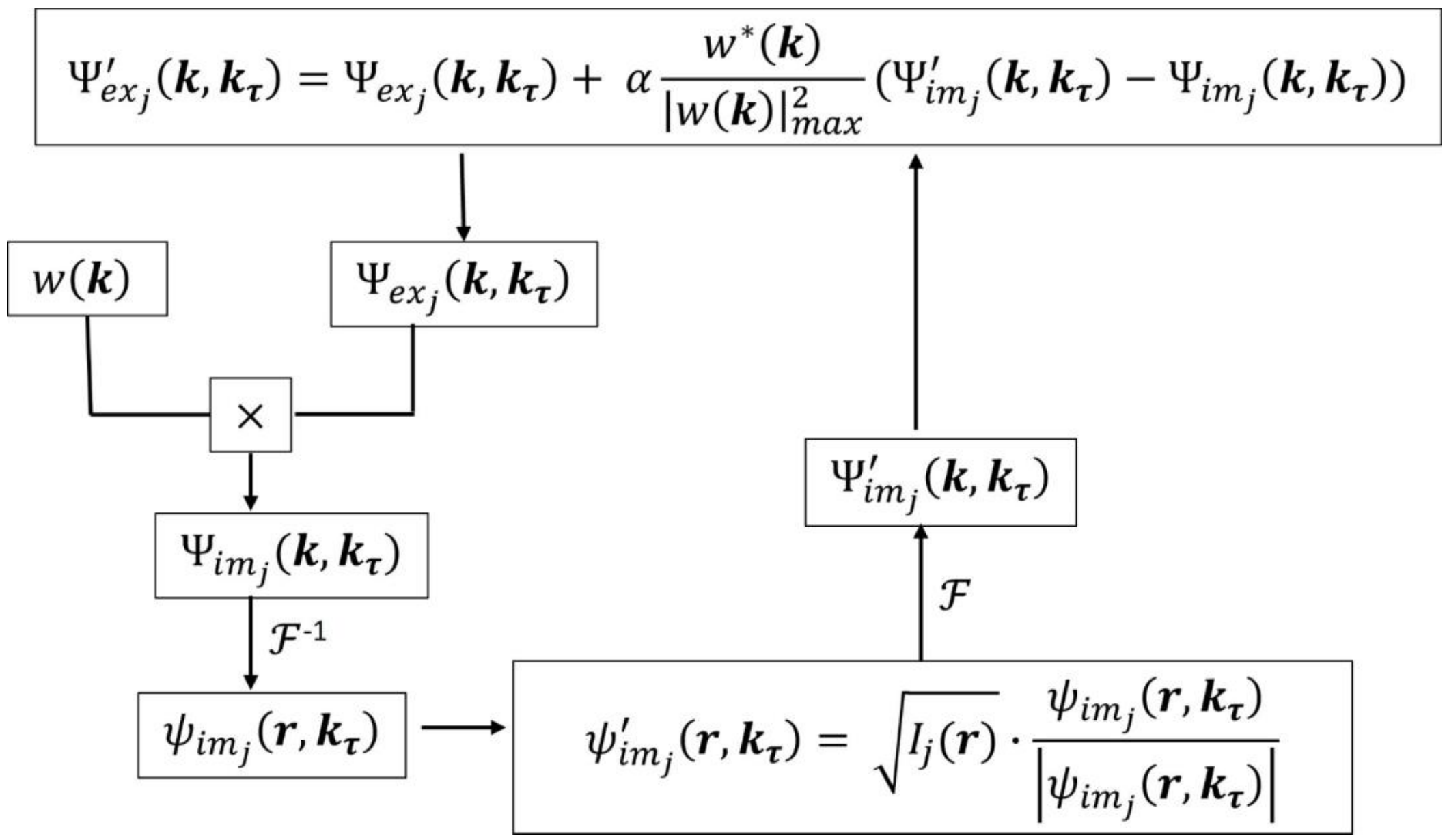

**Fig. S1. Flowchart of eFP exit wave reconstruction using the modified PIE algorithm.**

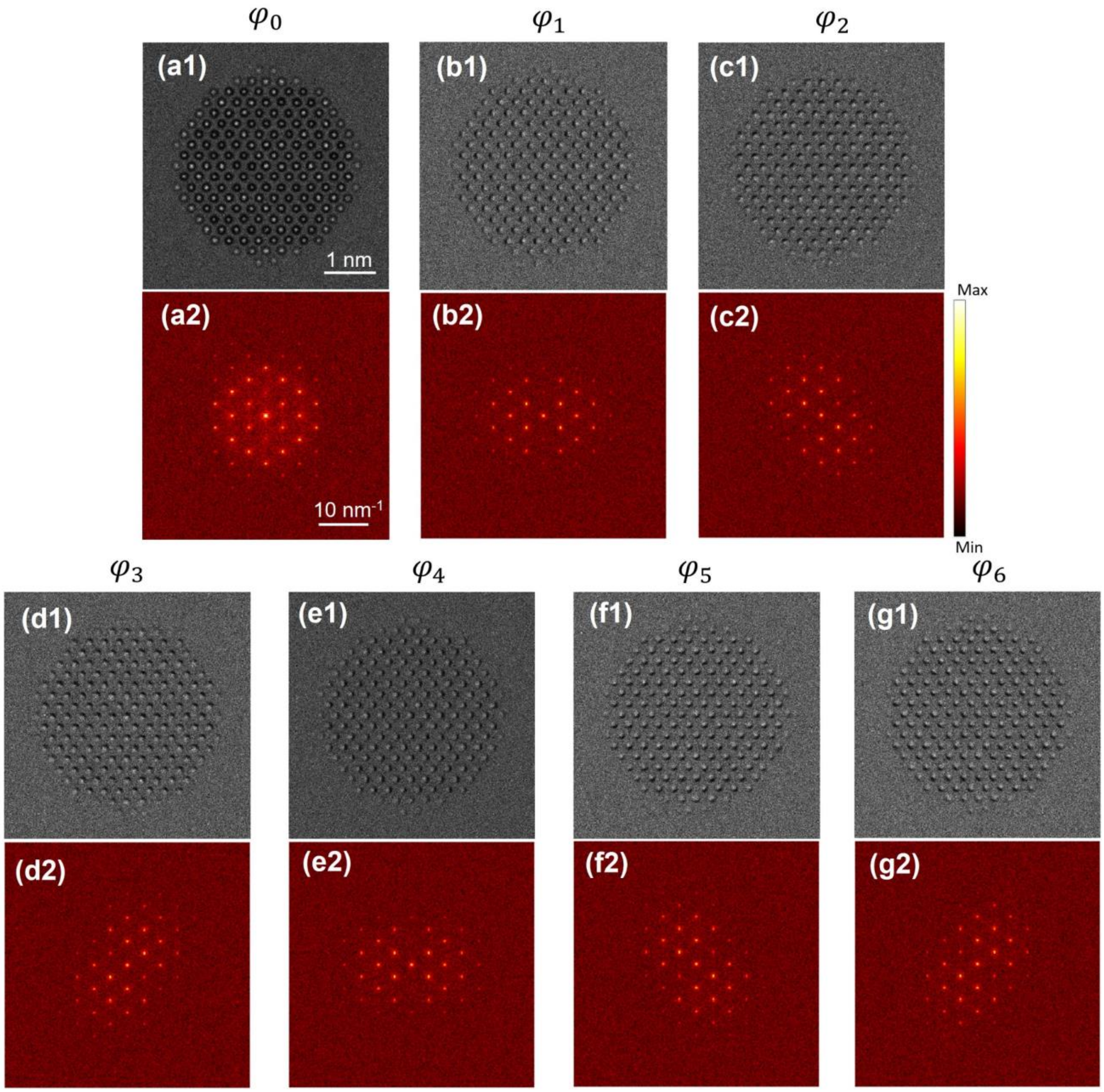

**Fig. S2. Simulated gold particle data at a beam tilt magnitude of 10.0 mrad. (a1)-(g1)** Axial and six tilted-illumination images with $\varphi_n$ indicating the illumination tilts; $\varphi_0$ corresponds to axial illumination, and $\varphi_1$- $\varphi_6$ to six tilted illuminations with evenly spaced azimuths and a constant tilt magnitude of 10.0 mrad. **(a2)-(g2)** Corresponding power spectra calculated from the amplitude of simulated image intensities. **(a1)-(g1)** are displayed with the scale bar shown in **(a1)**. Power spectra **(a2)-(g2)** are displayed with the scale bar shown in **(a2)**. The intensity of all the power spectra are weighted by a power of 0.2 for better visualisation of high-frequency information. The total fluence used was $4.6\times10^5$ $e^-/nm^2$.

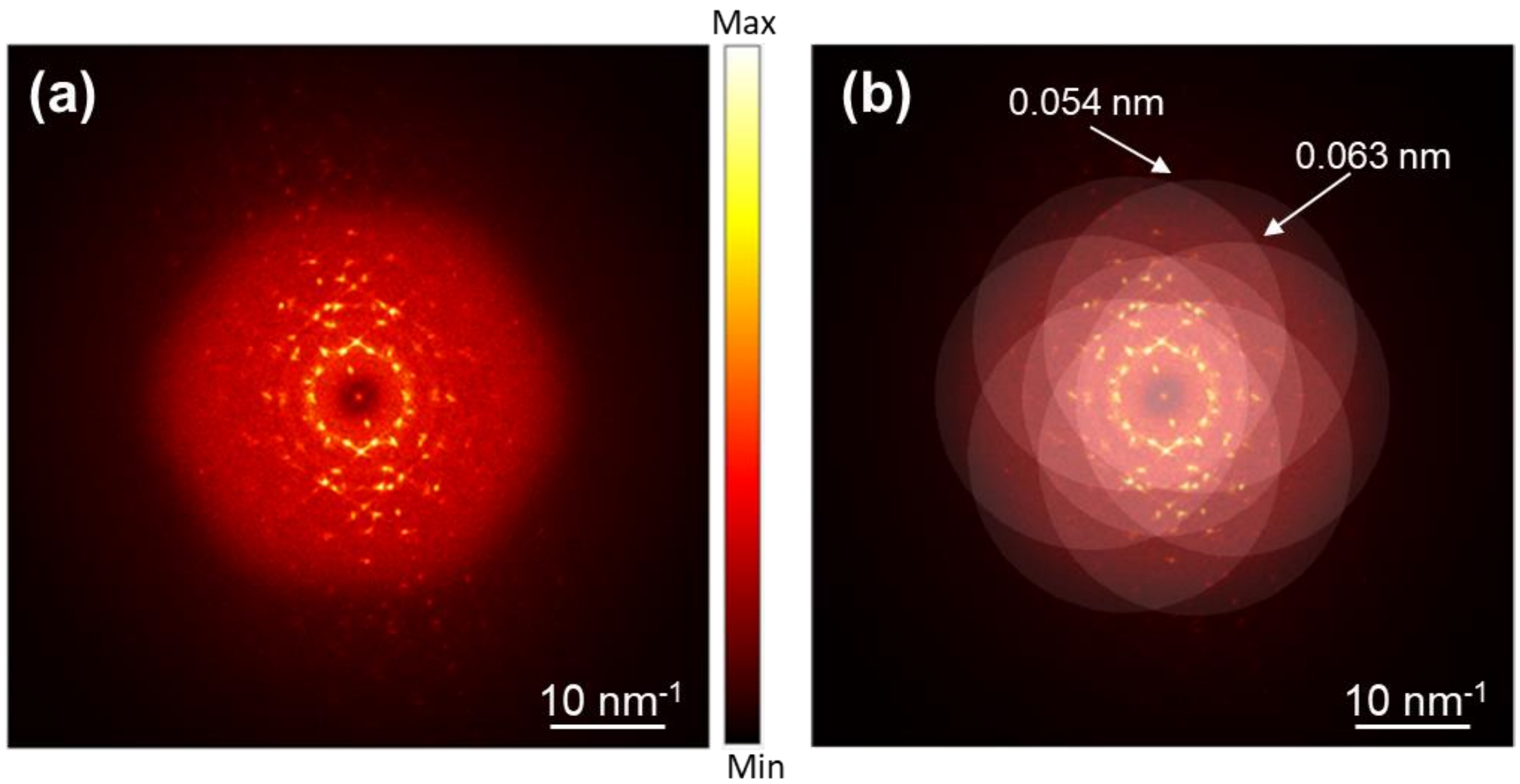

**Fig. S3. Power spectrum of the reconstructed phase and corresponding synthetic aperture.** **(a)** Power spectrum of reconstructed phase, as shown in Fig. 3 of the main text. **(b)** Spectrum with the corresponding synthetic aperture overlaid. The synthetic aperture is defined by a combination of the amplitude of axial wave transfer function and effective wave transfer function under tilted illuminations with a limit where the information transfer falls below 10%.

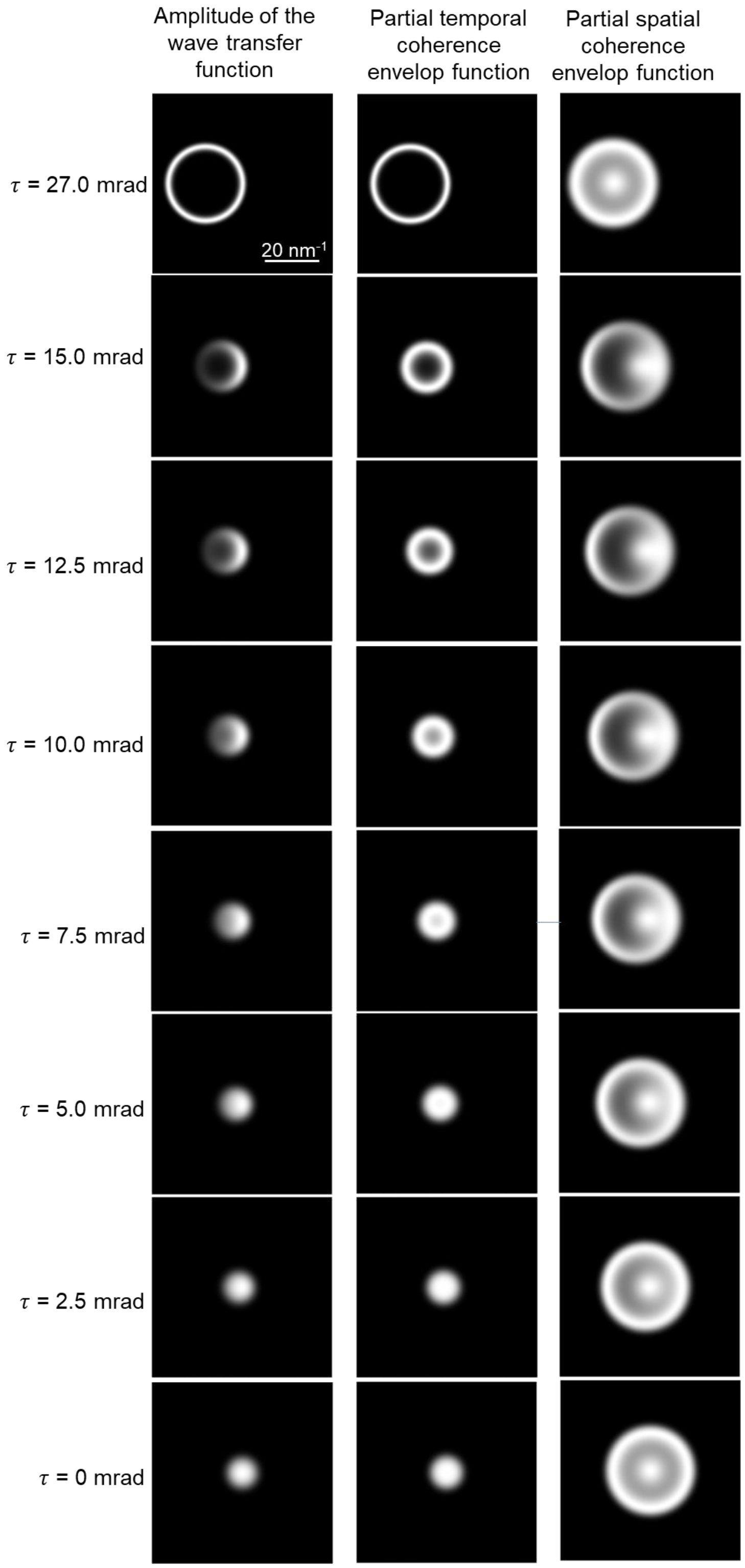

**Fig. S4. Effective wave transfer function and partial temporal and spatial coherence envelop functions calculated at different tilt magnitudes.** The calculations assume $C_1$ = -2000 nm and $C_3$ = 2.7 mm, with other parameters identical to those used in the apoferritin simulation described in the main text (**Table 1**). All images share the same scale bar shown in the top right image.

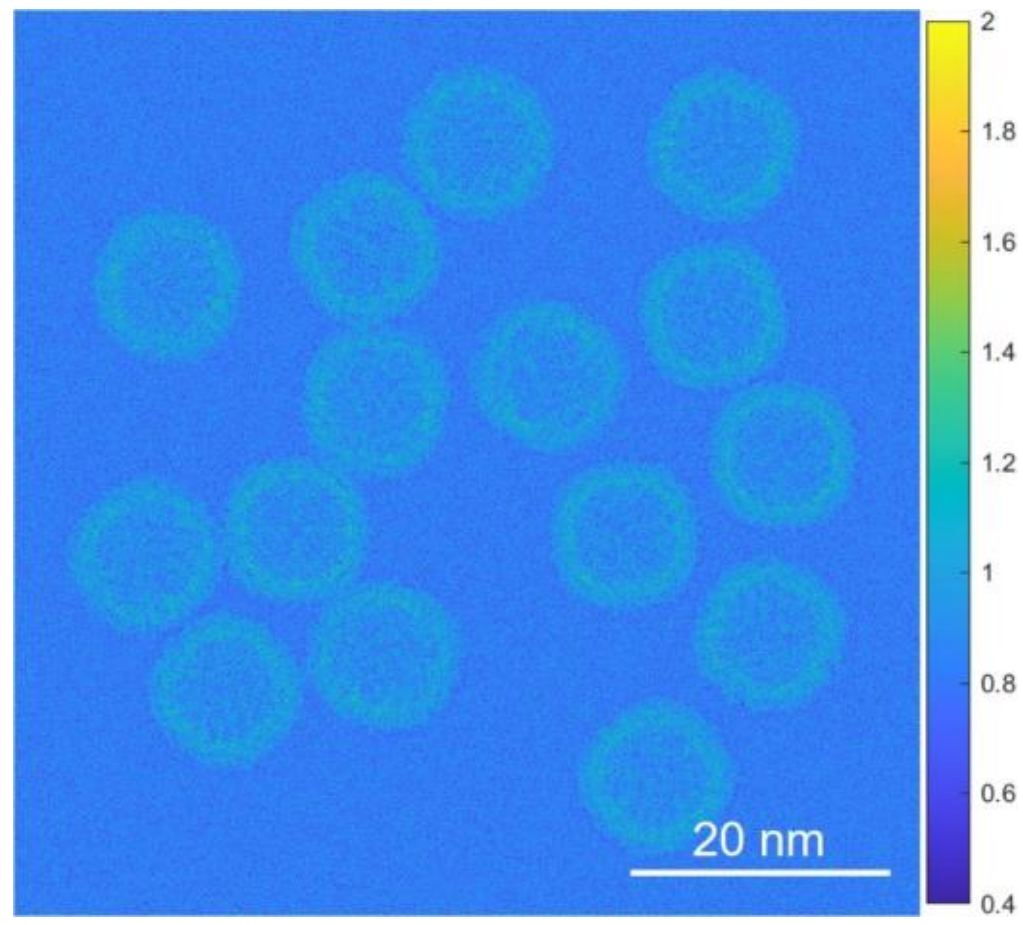

**Fig. S5. Simulated phase of apoferritin particles used as the ground truth for evaluating the quality of the reconstructed phase.** Phase scale in radians.

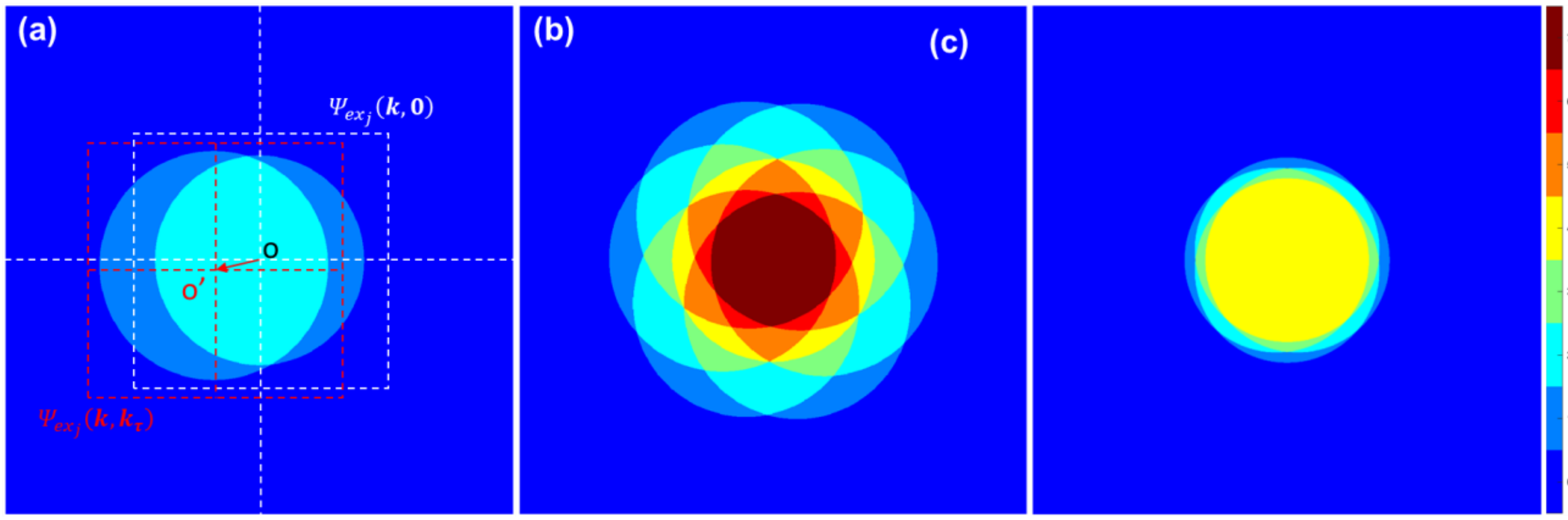

**Fig. S6. Schematic illustration of the synthetic aperture and measurement redundancy**. **(a)** Information corresponding to axial illumination (white dashed-square, $\Psi_{ex_j}(\boldsymbol{k},\boldsymbol{0})$) and that corresponding to a single tilted illumination image (red dashed-square, $\Psi_{ex_j}(\boldsymbol{k},\boldsymbol{k_\tau})$). **(b)** and **(c)** Synthetic apertures used in the reconstructions in **Fig. 3** and **Fig. 5** of the main text. Colour bar indicates measurement redundancy across different spatial frequency regions.

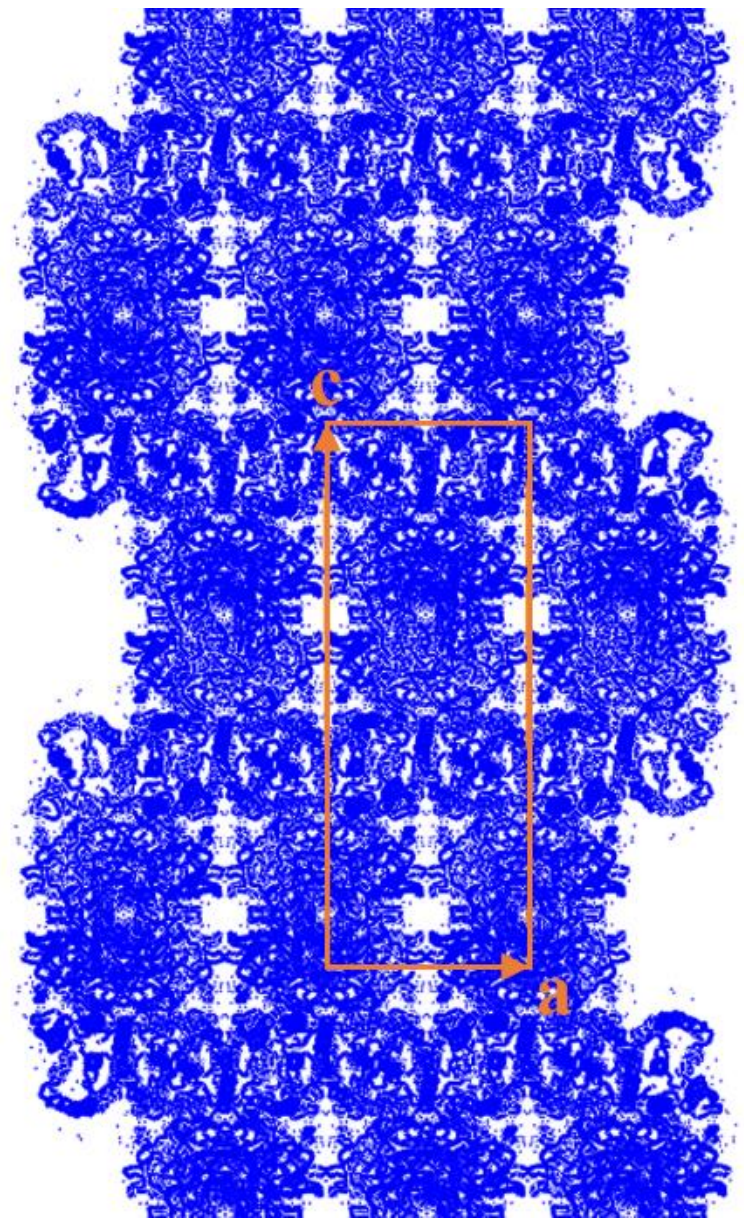

**Fig. S7. Cry11Aa crystal structure projected along [010].**

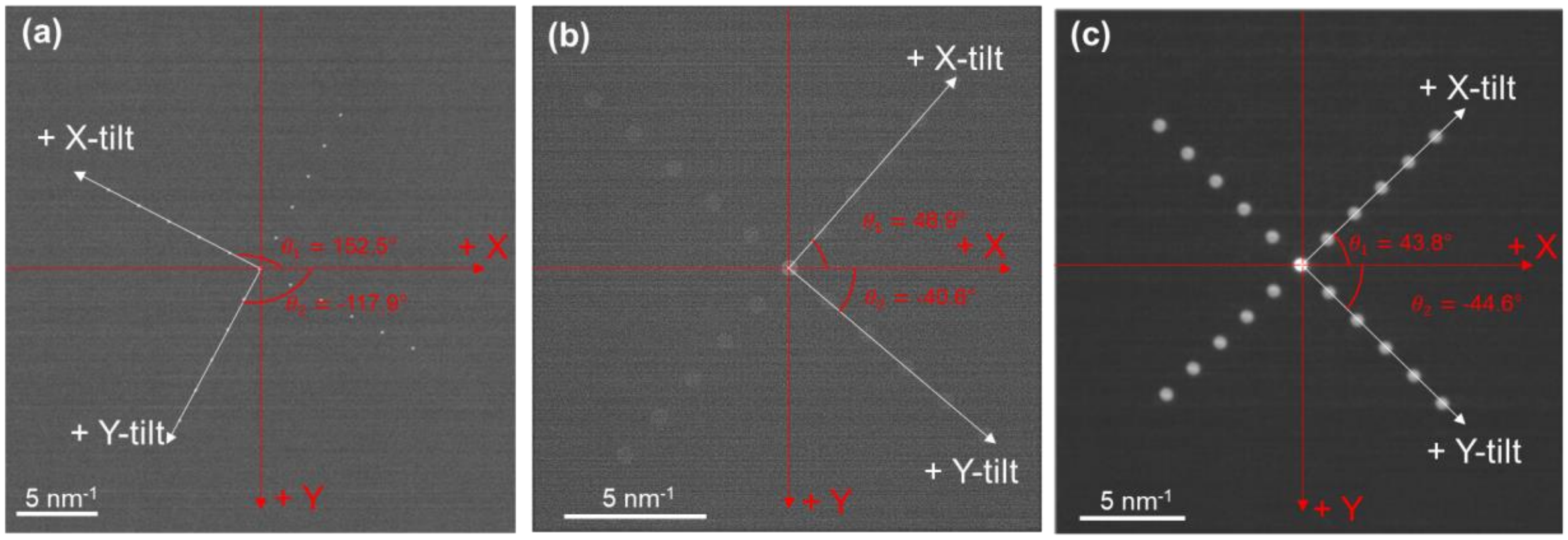

**Fig. S8. Tilt calibration for the experimental datasets. (a)** For the gold particle dataset collected on a JEM-ARM300F2. **(b)** For the rotavirus dataset collected on a JEM-Z300FSC. **(c)** For the Cry11Aa dataset collected on JEM-Z300FSC. In all cases the direct beam was defocused to protect the detector.

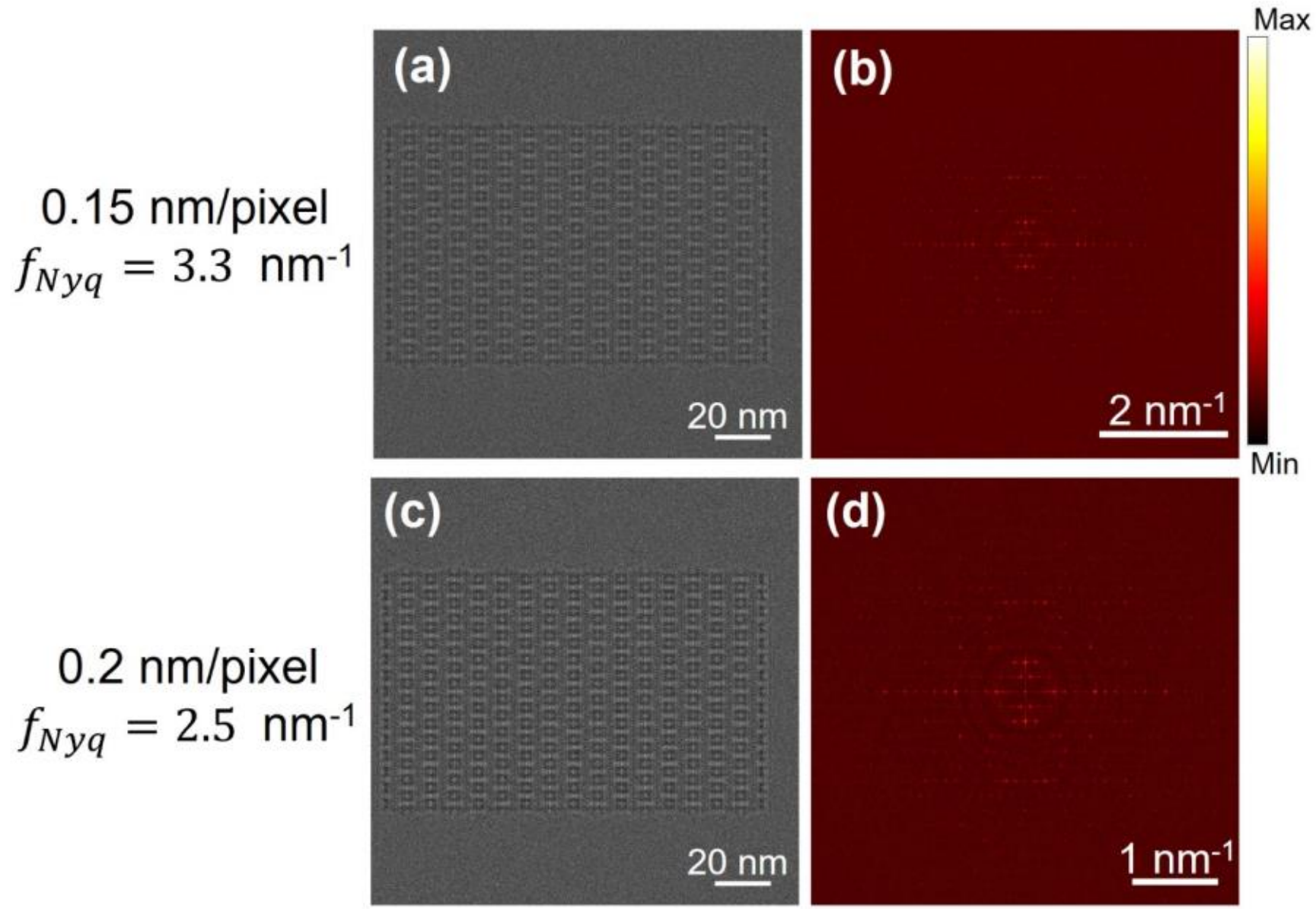

**Fig. S9. Simulated Crys11Aa data with different pixel samplings. (a)** and **(c)** tilted-illumination images (tilt magnitude 1.9 mrad) with a pixel sampling of 0.15 nm/pixel and 0.2 nm/pixel, respectively. **(b)** and **(d)** power spectra calculated from **(a)** and **(c)**. The total electron fluence is $3\times10^3$ $e^-/nm^2$. Data simulation details are as used in **Fig. 5** in the Main text, but with different pixel sampling and electron fluence. These represent two examples where the input eFP data fails to meet the Nyquist sampling criterion.

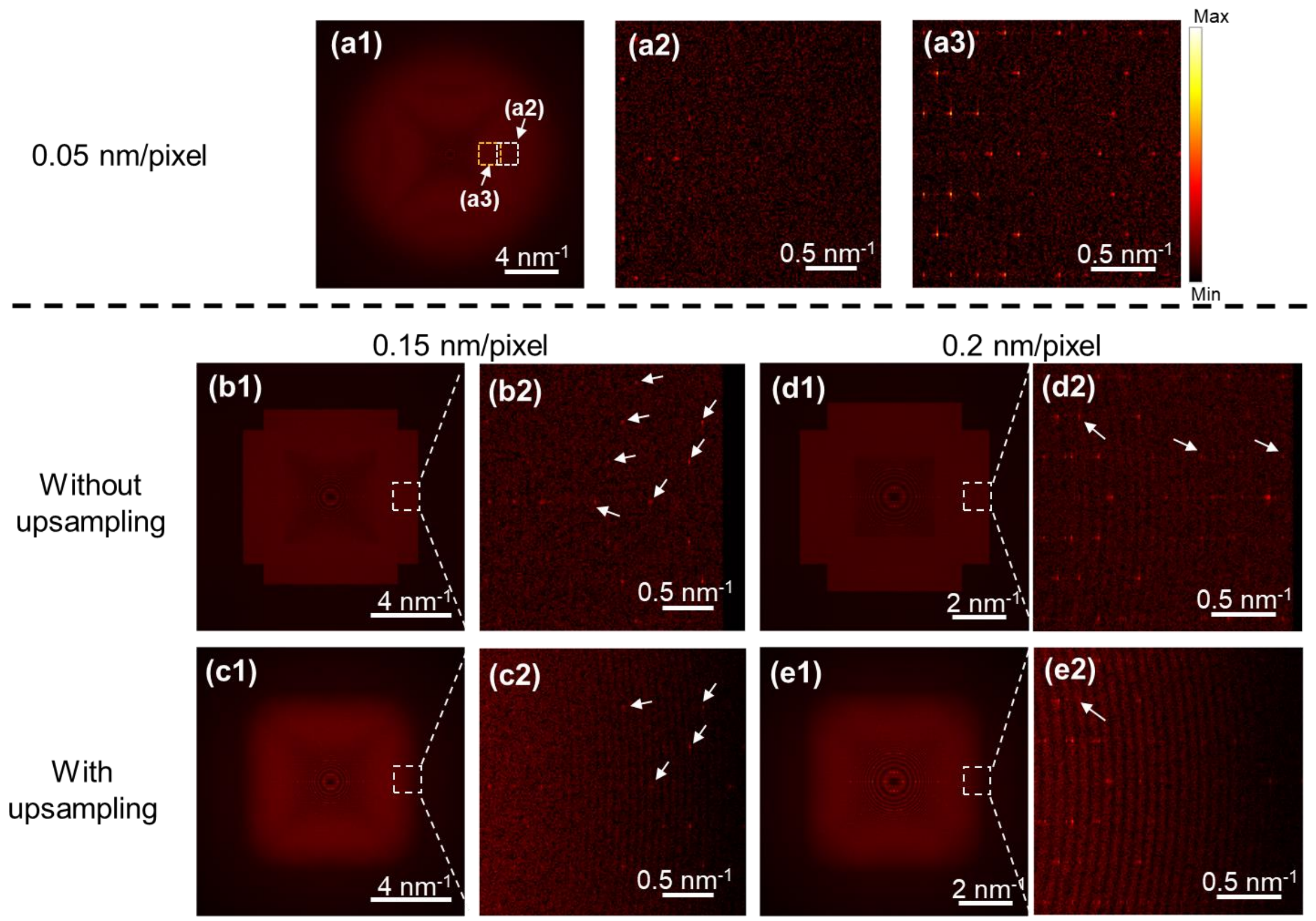

**Fig. S10. Upsampling for anti-aliasing. (a1)** Power spectrum calculated from the recovered phase of the simulated Cry11Aa dataset with a pixel sampling of 0.05 nm/pixel. **(a2)** and **(a3)** Enlarged views of the marked regions in (**a1**). **(b1)** Power spectrum calculated from the recovered phase with a pixel sampling of 0.15 nm/pixel for reconstruction without upsampling. **(b2)** Enlarged region marked in **(b1)**. **(c1)** and **(c2)** as for **(b1)** and **(b2)**, but with upsampling. **(d1)** Power spectrum calculated from the recovered phase from the dataset with a pixel sampling of 0.2 nm/pixel without upsampling. **(d2)** Enlarged region marked in **(d1)**. **(e1)** and **(e2)** as for **(d1)** and **(d2)**, but with upsampling. **(a1)-(a3)** are used as a comparison with **(b1)-(e2)**. **(a2)** and **(a3)** correspond to the same frequency regions as those shown in **(b2)-(c2)** and **(d2)-(e2)**, respectively. The arrow marks artefacts due to aliasing.

**Table S1. Corrected axial aberration coefficients to third order in the wave aberration function**

| Aberrations | Amplitude | Azimuth (degree) |
|---|---|---|
| Defocus, $C_1$ (nm) | 5 | |
| Two-fold astigmatism, $A_1$ (nm) | 2.0 | -89.8 |
| Three-fold astigmatism, *A2 (nm)* | 30.1 | -11.0 |
| Axial coma, $B_2$ (nm) | 22.4 | -41.9 |
| Four-fold astigmatism, *A3* (nm) | 162 | -19.5 |
| Axial star aberration, *S3 (nm)* | 1256 | -27.5 |
| Spherical aberration, $C_3$ (mm) | 0.0015 | |

**Table S2. Beam tilt calibration**

| Dataset | Tilt unit (mrad/143 bytes) | $\theta_1$* (degree) | $\theta_2$* (degree) |
|---|---|---|---|
| Gold particle | 0.9 | 152.5 | -117.9 |
| Rotavirus | 3.8 | 48.9 | -40.6 |
| Cry11Aa crystal | 3.8 | 43.8 | -44.6 |

*$\boldsymbol{\theta_1}$ is defined as the angle between the image x-axis and the beam + x-tilt; $\boldsymbol{\theta_2}$ is defined as the angle between the image x-axis and the beam + y-tilt. This definition is illustrated in **Fig. S5**.